\documentclass[structabstract]{aa}  

\usepackage{graphicx}
\usepackage{subfig}
\usepackage[round]{natbib}
\usepackage{color}
\usepackage{amsmath}
\usepackage{txfonts}

\newcommand{\Phib}{\Phi_{\mathrm{b}}}
\newcommand{\Omegab}{\Omega_{\mathrm{b}}}

\newcommand{\RCR}{R_{\mathrm{CR}}}
\newcommand{\ROLR}{R_{\mathrm{OLR}}}

\newcommand{\Kpc}{\ \mathrm{kpc}}

\newcommand{\Gyr}{\ \mathrm{Gyr}}
\newcommand{\Myr}{\ \mathrm{Myr}}
\newcommand{\kmsec}{\ \mathrm{km}\ \mathrm{s}^{-1}}
\newcommand{\kmseckpc}{\ \mathrm{km}\ \mathrm{s}^{-1}\ \mathrm{kpc}^{-1}}
\newcommand{\Mh}{M_{\mathrm{h}}}
\newcommand{\Mb}{M_{\mathrm{b}}}
\newcommand{\Msun}{M_{\odot}}

\newcommand{\ab}{a_{\mathrm{b}}}
\newcommand{\athin}{a_{\mathrm{thin}}}
\newcommand{\athick}{a_{\mathrm{thick}}}
\newcommand{\bthin}{b_{\mathrm{thin}}}
\newcommand{\bthick}{b_{\mathrm{thick}}}

\newcommand{\rhodi}{\rho_{\mathrm{d},i}}
\newcommand{\Phidi}{\Phi_{\mathrm{d},i}}
\newcommand{\sigmaRi}{\sigma_{R,i}}
\newcommand{\sigmazi}{\sigma_{z,i}}
\newcommand{\sigmaphii}{\sigma_{\phi,i}}

\newcommand{\Phih}{\Phi_{\mathrm{h}}}
\newcommand{\ah}{a_{\mathrm{h}}}
\newcommand{\vc}{v_{\mathrm{c}}}
\newcommand{\Mbar}{M_{\mathrm{bar}}}
\newcommand{\as}{a_{\mathrm{s}}}
\newcommand{\de}{\mathrm{d}}

\newcommand{\vai}{v_{\mathrm{a},i}}

\newcommand{\Tbar}{T_\mathrm{bar}}

\newcommand{\Rg}{R_\mathrm{g}}
\newcommand{\Rh}{R_\mathrm{h}}

\newcommand{\zthin}{z_\mathrm{thin}}
\newcommand{\zthick}{z_\mathrm{thick}}
\newcommand{\Mthin}{M_\mathrm{thin}}
\newcommand{\Mthick}{M_\mathrm{thick}}
\newcommand{\Roto}{R_{-1:1}}

\newcommand{\Rcor}{R_\mathrm{COR}}
\newcommand{\vphiOLR}{v_{\phi,\mathrm{OLR}}}
\newcommand{\vphioto}{v_{\phi,\mathrm{-1:1}}}
\newcommand{\vphitto}{v_{\phi,\mathrm{-3:1}}}
\newcommand{\vphifto}{v_{\phi,\mathrm{-4:1}}}
\newcommand{\rvol}{R_\mathrm{vol}}
\newcommand{\hvol}{h_\mathrm{vol}}
\newcommand{\lv}{l_\mathrm{v}}
\newcommand{\RNum}[1]{\uppercase\expandafter{\romannumeral #1\relax}}
\newcommand{\bfx}{\mathbf{x}}
\newcommand{\bfv}{\mathbf{v}}
\newcommand{\bfxo}{\mathbf{x}_0}
\newcommand{\bfvo}{\mathbf{v}_0}
\newcommand{\tf}{t_\mathrm{f}}

\begin{document}
\title{3D test particle simulations of the Galactic disks. The
       kinematical effects of the bar.}

\author{G. Monari,
  T. Antoja
  \and
  A. Helmi
}

\institute{Kapteyn Astronomical Institute, Rijksuniversiteit Groningen, 
  P.O. Box 800, 9700 AV Groningen, The Netherlands\\
  \email{monari@astro.rug.nl}
}

\abstract
{}
{To study the imprints of a rotating bar on the kinematics of stars in
  the thin and thick disks throughout the Galaxy.}
{We perform test particle numerical simulations of the thin and thick
  disks in a 3D Galactic potential that includes a halo, a bulge, thin
  and thick disks, and a Ferrers bar.  We analyze the resulting
  velocity distributions of populations corresponding to both disks,
  for different positions in the Galaxy and for different structural
  parameters of the bar.
}
{We find that the velocity distributions of the disks are affected by
  the bar, and that strong transient effects are present for
  approximately 10 bar rotations after this is introduced
  adiabatically. On long (more realistic) timescales, the effects of
  the bar are strong on the kinematics of thin disk stars, and weaker
  on those in the thick disk, but in any case significant.
  Furthermore, we find that it is possible to trace the imprints of
  the bar also vertically and at least up to $z\sim1\Kpc$ for the thin
  disk and $z\sim 2 \Kpc$ for the thick disk.  }
{}

	
\maketitle

\section{Introduction} \label{sect:Intro} The study of kinematic
groups in the Solar Neighbourhood dates back to the work of
\cite{Proctor1869} and \cite{Kapteyn1905}, with the discovery of the
Hyades and Ursa Major groups. In modern times, the analysis of
\emph{Hipparcos} data and other surveys (\citealt{Dehnen1998};
\citealt{Famaey2005, Antoja2008}) revealed the presence of many other
local kinematic structures with varying properties, some of which are
composed by stars of a wide range of age.  More recently,
\cite{Antoja2012} using \emph{RAVE} data demonstrated that some of the
known local kinematical groups extend beyond the Solar neighbourhood
at least as far as $\sim 1\Kpc$ from the Sun on the plane, and $\sim
0.7\Kpc$ below it.

These main kinematic groups are in the thin disk. Their populations
spread (age and metallicity) indicates that they are unlikely to be
remnants of disrupted clusters (\citealt{Eggen1996}). However, they
can be explained in terms of the influence of the non-axisymmetric
components of the Galaxy (bar and spiral arms) on the orbits of stars
(\citealt{Dehnen2000, Fux2001, DeSimone2004, Quillen2005, Antoja2009,
  Antoja2011}).  For instance, the existence of the Hercules stream, a
local group of stars moving outwards in the disk and lagging the
circular rotation, has been related to the bar's Outer Lindblad
Resonance (\citealt{Dehnen2000}).

On the other hand, much of the past work on the kinematics of the
thick disk has explored the presence of phase-space substructure due
to minor mergers and accretion events (\citealt{Helmi2006,
  Villalobos2008, Villalobos2009}).  For example it has been suggested
that the Arcturus stream (\citealt{Eggen1971}) may have an
extra-Galactic origin (\citealt{Navarro2004}). However, it has also
been advocated that this stream could have a Galactic origin
(\citealt{Williams2009, Minchev2009}), and be a signature of a bar's
resonance (\citealt{Antoja2009, GardnerFlynn2010}).  Yet this proposal
may be considered somewhat speculative, as it is based on simulations
of the kinematics on the plane of the Galaxy, that is ignoring the
vertical motion, and it is a priori not so clear whether a resonance
could affect stars of the thick disk that spend so much time far from
the Galactic plane.

We thus currently have a limited understanding as to whether the
non-axisymmetric components (bar and spiral arms) can induce kinematic
structure in the thick disk, and how to distinguish this from the
substructures associated to accretion events.  It is only recently
that \cite{Solway2012}, using controlled N-Body simulations to study
radial migration, quantified that transient spiral arms can change the
angular momentum of stars in the thick disk almost as much as they do
in the thin disk.
 
This Introduction serves to present the main motivation of our paper,
namely to study the influence of the Galactic bar on the thin and
thick disk in 3D, paying special attention to the vertical dimension.
Our final aim is to establish if kinematic groups or other signatures
of the bar may be present far from the Galactic plane, that is for
orbits with large vertical oscillations, both in the thin and thick
disk.  To this end, we perform controlled test particle orbit
integrations in a Galactic potential that includes a bar.  Our
simulations present a number of improvements with respect to previous
studies, mainly that the integrations are done in a potential that is
3D, and that we study both the thin and thick disks.

In Section \ref{sect:simu} we give details of the simulation
techniques and choice of initial conditions.  In Section
\ref{sect:Results} we analyze the resulting velocity distributions of
the thin and thick disks in localized volumes around the Solar
Neighbourhood, both in the Galactic plane as well as away from it.
Finally, in Section \ref{sect:con} we present the implications of our
results and our conclusions.

\section{Simulation methods}\label{sect:simu}

Our approach to study the characteristics of the velocity distribution
across the Galaxy is based on a ``forward integration
technique''. First we make discrete realizations of specific
distribution functions, then we integrate these initial conditions
forward in time, using an adaptive sizestep Bulirsch-Stoer
(\citealt{NumericalRecipes}).  Finally we analyze the resulting coarse
grained distribution inside finite volumes.

This approach may be contrasted to the ``backward integration
technique'' (e.g., \citealt{Dehnen2000, GardnerFlynn2010})
\footnote{Starting from a time $t=\tf$, a point in space $\bfx$ and a
  grid of velocities $\bfv^j$, the orbits are integrated backward in
  time.  At the end of each integration, each orbit reaches a point
  $(\bfxo^j,\bfvo^j)$, associated with the value of an analytic
  phase-space distribution function $f(\bfxo^j,\bfvo^j,t=t_0)$
  corresponding to the unbarred potential.  The collisionless
  Boltzmann equation states that the fine grained distribution
  function remains constant along the orbits. It is then possible to
  associate to each point of the grid of velocities a value of the
  phase-space density, since
  $f(\bfx,\bfv^j,t=\tf)=f(\bfxo^j,\bfvo^j,t=t_0)$, and to see how the
  velocity space is populated at the point $\bfx$ of the Galaxy.},
where the velocity distribution at a single point in configuration
space is obtained.  Our method instead requires a large number of
particles and is computationally more expensive, but it may be
considered more realistic in reproducing how the phase-space is
populated and is possibly more appropriate for comparison to the
observations, which always probe a finite volume.

All the particles of our simulations are integrated in a potential
composed by two parts: an axisymmetric part (formed by a bulge, a thin
disk, a thick disk and a dark halo) and a non-axisymmetric part (the
bar).  The bar is introduced in the potential gradually with time and
replaces the bulge.  The mass of the bulge is continuously transferred
to the bar, and becomes null when the bar has grown completely (see
details in Section \ref{sect:bar})

Here and in the rest of the paper $r$ and $R$ are the Galactocentric
spherical and cylindrical radial coordinate respectively, $\phi$ is
the polar angle measured from the long axis of the bar in the
direction of Galactic rotation, and $z$ the vertical
coordinate. Moreover, $v_R\equiv-\dot{R}$ is the radial Galactocentric
velocity (positive towards the center of the Galaxy), $v_\phi\equiv
R\dot{\phi}$ is the azimuthal velocity (positive in the direction of
the Galactic rotation) and $v_z\equiv\dot{z}$.

\subsection{Axisymmetric part of the potential}\label{sect:axi}
\begin{table}
  \caption{Parameters of the two axisymmetric models A0+B1
    and A0+B2.}
  \label{tab:potaxi}
  \centering
  \begin{center}
    \begin{tabular}{ll} \hline \hline Parameter & \\ \hline
      $\Mb(\Msun)$ & $1\times 10^{10}$ (B1)\\
      & $2\times 10^{10}$ (B2)\\
      $\ab(\Kpc)$ & $1$ \\
      $\Mh(\Msun)$ & $8\times 10^{11}$ \\
      $\ah(\Kpc)$ & $15.84$ \\
      $c$ & $18$ \\
      $\Mthin(\Msun)$ & $5.76\times 10^{10}$ \\
      $\Mthick(\Msun)$ & $1.14\times 10^{10}$ \\
      $\athin(\Kpc)$ & $3.30$ \\
      $\athick(\Kpc)$ & $3.05$ \\
      $\bthin(\Kpc)$ & $0.13$ \\
      $\bthick(\Kpc)$ & $0.98$ \\\hline
    \end{tabular}
  \end{center}
\end{table}

Below we give a detailed description of the individual components
contributing to the axisymmetric part of the potential. Table
\ref{tab:potaxi} lists their characteristic parameters.

The bulge follows a \cite{Hernquist1990} potential,
\begin{equation}
  \Phib(r)=-\frac{G\Mb}{\ab+r},
\end{equation}
with $\ab=1\Kpc$.
We choose two different values for the mass of the bulge (and,
therefore, of the bar): $\Mb=10^{10}\Msun$ (B1) and
$\Mb=2\times10^{10}\Msun$ (B2, see Section \ref{sect:bar}).

We represent the thin and thick disk potentials with
\cite{MiyamotoNagai1975} models,
\begin{equation}\label{eq:MNpot}
  \Phidi(R,z)=-\frac{GM_i}{\sqrt{R^2+\left(a_i+\sqrt{z^2+b_i^2}\right)^2}},
\end{equation} 
where ``$i$'' stands for ``thin'' or ``thick''.
The mass ratio between the two disks is $20\%$ and this gives a
thick-to-thin disk density normalization of $\sim 10\%$ near the Sun,
similarly to what is observed in our Galaxy \citep{Juric2008}.  We
have chosen the Miyamoto-Nagai functional form to represent the disks
because of their mathematical simplicity which results in
computational convenience. Since in reality the Galactic disks follow
more closely an exponential form in the density (\citealt{BT2008}), we
set the characteristic parameters of the Miyamoto-Nagai model to
resemble the {\it gravitational potential} of two exponential disks
with radial scale lengths $R_h = 3\Kpc$ (for both disks) and vertical
scale lengths $\zthin = 0.3\Kpc$ and $\zthick = 1\Kpc$.  Although the
potentials are similar, the forces differ. At $R = 8 \Kpc$ the
Miyamoto-Nagai thin disk overestimates the exponential disk vertical
force $f_z$ up to $30\%$ at $\left| z \right| \lesssim 0.3 \Kpc$,
while it underestimates the force by $\sim 15\%$ at $|z| \gtrsim
0.3\Kpc$.  The exponential thick disk force in the Miyamoto-Nagai
model is underestimated $\sim 10-15\%$ at all heights we
consider. These small differences in the force field can presumabxly
be regarded as being smaller than the uncertainties in the true form
and values of the characteristic parameters of the Milky Way
gravitational field.

The dark halo follows the NFW potential (\citealt{NFW})
\begin{equation}
  \Phih(r)=-\frac{G\Mh}{\ln(1+c)-c/(1+c)}\frac{\ln(1+r/\ah)}{r},
\end{equation}
with $\Mh=8\times10^{11}\Msun$, $\ah = 15.84$ and $c=18$
(\citealt{Battaglia2005}).

The axisymmetric potential consisting of halo, thin and thick disks is
referred as ``A0'' (and is the same for all models explored), while
when the bulge is added we refer to A0+B1 or A0+B2 depending on the
mass of the bulge.

\subsection{Bar potential}\label{sect:bar}
\begin{table}
  \caption{Parameters of the bar and location of the main resonances.}
  \label{tab:bar}
  \centering
  \begin{center}
    \begin{tabular}{@{\extracolsep{-5pt}}lll}\hline\hline
      Parameter     & GB     & LB\\ \hline
      % $\Mbar/\Msun$ & $10^{10}$ (GB1), $2\times 10^{10}$ (GB2) & $10^{10}$ (LB1),  $2\times 10^{10}$ (LB2) \\
      $\Mbar(\Msun)$   & $1\times10^{10}$ (GB1) & $1\times10^{10}$ (LB1)\\
      & $2\times 10^{10}$ (GB2)& $2\times 10^{10}$ (LB2) \\
      % $\Mbar/\Msun$   & GB1: $10^{10}$         & LB1: $10^{10}$\\
      % & GB2: $2\times 10^{10}$ & LB2: $2\times 10^{10}$  \\
      $a(\Kpc)$       & $3.5$  & $3.9$ \\
      $b(\Kpc)$      & $1.4$  & $0.6$ \\
      $c(\Kpc)$      & $1.0$  & $0.1$ \\
      $\RCR(\Kpc)$   & $4.54$ (A0+GB1) & $4.57$ (A0+LB1) \\
      & $4.91$ (A0+GB2) & $4.94$ (A0+LB2) \\
      $\ROLR(\Kpc)$  & $7.40$ (A0+GB1) & $7.40$ (A0+LB1) \\
      & $7.69$ (A0+GB2) & $7.69$ (A0+LB2) \\
      $\Roto(\Kpc)$  & $9.91$ (A0+GB1) & $9.91$ (A0+LB1) \\
      & $10.22$ (A0+GB2) & $10.22$ (A0+LB2) \\\hline
    \end{tabular}
  \end{center}
 
\end{table}
The 3D bar used in our simulations is a \cite{Ferrers1870} bar,
whose density profile is given by
\begin{equation}\label{eq:ferrers}
  \rho_{\mathrm{bar}}(m^2) = \left\{ 
    \begin{array}{l l}
      \rho_0(1-m^2/a^2)^n & \quad m \leq a\\
      0 & \quad m > a,
    \end{array} \right.
\end{equation}
where $m^2\equiv a^2 \left(x^2/a^2+y^2/b^2+z^2/c^2\right)$.  We choose
$n=2$ and explore two sets of axis lengths taken from
\cite{BissantzGerhard2002} and \cite{LopezC2007}, respectively. We dub
them ``Galactic Bar'' (GB) and ``Long Bar'' (LB), in the same fashion
as \cite{GardnerFlynn2010}.  The bar parameters are listed in Table
\ref{tab:bar}.

For the pattern speed of the bar we use the value $\Omegab = 50
\kmseckpc$, i.e. in the range of current determinations, which is
between $35$ and $60\kmseckpc$ (\citealt{Gerhard2011}). It is
important to notice that the effects on the velocity distribution in a
certain volume are mainly set by the proximity to the main resonances.
This implies that a faster (slower) pattern speed yields similar
effects as the original bar in outer (inner) volumes.

As anticipated, the bar potential grows adiabatically in the
simulations. Specifically, each simulation starts with an axisymmetric
potential (either A0+B1 or A0+B2), and the mass of the bulge is
transferred to the bar using the function
\begin{equation}\label{eq:growth}
  \eta(t)=\left(\frac{3}{16}\xi^5-\frac{5}{8}\xi^3+\frac{15}{16}\xi+\frac{1}{2}\right),\quad \xi=2\frac{t}{t_1}-1,
\end{equation}
which runs smoothly from $\eta(0)=0$ to $\eta(t_1)=1$
(\citealt{Dehnen2000}).  In particular:
\begin{itemize}
\item for $t\in[0,t_1]$, $\Mbar(t)=\eta(t)\Mb(0)$ and
  $\Mb(t)=[1-\eta(t)]\Mb(0)$,
\item for $t>t_1$, $\Mbar(t)=\Mb(0)$ and $\Mb(t)=0$.
\end{itemize}
In our models $\Mb(0)=\Mbar(t\geq t_1)=10^{10}\Msun$ (B1) and
$\Mb(0)=\Mbar(t\geq t_1)=2\times 10^{10}\Msun$ (B2), which is
consistent with the mass estimates by \cite{Dwek1995} and
\cite{Zhao1996}.

In all our simulations the bar grows following Eq. (\ref{eq:growth}),
up to time $t_1$.  The choice of $t_1$ is not particularly relevant,
as it was shown in \cite{Minchev2010} that a longer growth time of the
bar only linearly delays the effects of the bar (in this case, the
formation of the kinematic structures).  Throughout this paper we
always use $t_1=2\Tbar\simeq 246\Myr$ as in \cite{Fux2001}, where
$\Tbar=2\pi/\Omegab$.

\begin{figure}
  \centering
  \includegraphics[width=\columnwidth]{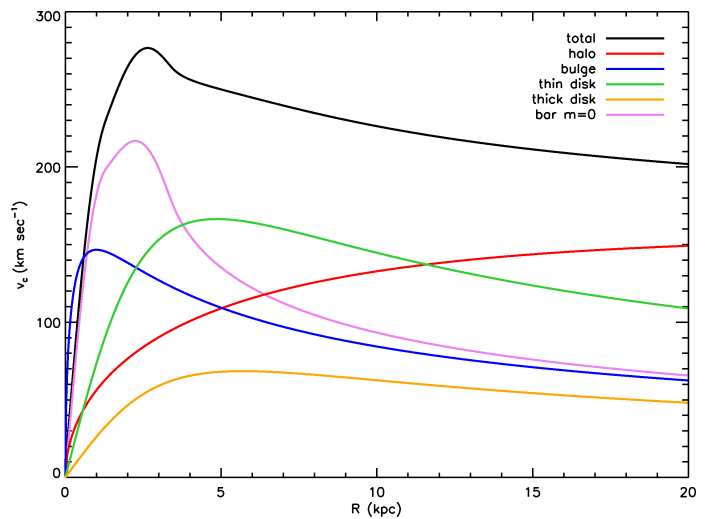}
  \caption{Circular velocity curves of the components of A0 (halo,
    thin and thick disk), bulge B2 and the $m=0$ Fourier component of
    the bar GB2. The solid black line is the resulting circular
    velocity for model A0+GB2.  }\label{fig:vc}
\end{figure}

In Fig. \ref{fig:vc} we show the circular velocity curves of the
components of A0 and of the bulge B2.  In the same figure we plot the
circular velocity of the $m=0$ term of the Fourier decomposition of
the bar potential GB2 (GB with $\Mbar=2\times 10^{10}\Msun$). Note
that, although the total mass of the bulge and the bar are the same,
inside $R=20\Kpc$ their velocity curves (blue and purple lines,
respectively) differ because the enclosed masses are different.  While
the Hernquist bulge extends to infinity, all the bar's mass is
confined inside a radius equal to its semi-major axis (see
Eq. \ref{eq:ferrers}).  As a consequence, when the bar growth has been
completed, the resulting circular velocity curve has changed.  For
this reason, for our analysis we use the circular velocity curve
$\vc(R)$ obtained by adding the contributions of the dark halo, thin
and thick disk and $m=0$ component of the bar potential (A0+GB2).
This curve is represented by the black line in Fig. \ref{fig:vc}.  We
see how the total rotation curve is mostly influenced by the bar for
approximately $R < 5\Kpc$, by the thin disk for $5\Kpc<R<10\Kpc$ and
by the dark halo for $R>10\Kpc$.  The value of the circular velocity
at Solar radius, and for $z = 0$ is $\vc(8\Kpc)\simeq 222\kmsec$ for
A0+GB1 and $\vc(8\Kpc)\simeq 234\kmsec$ for A0+GB2.

\subsection{Initial conditions}\label{sect:IC}
We use two sets of initial conditions for our simulations: ICTHIN and
ICTHICK.  ICTHIN mimics the typical kinematics and density
distribution of the intermediate age population of the thin disk,
while ICTHICK represents the thick disk. We generate low (LR) and high
resolution (HR) realizations, that have $N=5\times10^7$ and $N=10^9$
particles respectively. We perform HR simulations for our standard
Galactic model (A0+GB2, see below), to help to distinguish the
resonant features that are in the wings of the distribution and often
hidden in the noise.

The positions $(R,\phi,z)$ of the particles in both disks are
distributed following Miyamoto-Nagai densities, corresponding to the
potentials described in Section \ref{sect:axi}.  The density
distribution of each disk, $\rhodi$, is derived from
Eq. (\ref{eq:MNpot}) through Poisson's equation
(for the complete expression see \citealt{MiyamotoNagai1975}).  The
$(R,z)$ positions of the particles are randomly drawn from these
profiles with a method based on the Von Neumann's rejection technique
(\citealt{NumericalRecipes}) and $\varphi$ is generated uniformly
between $0$ and $2\pi$.

Once the positions are generated, the velocities are assigned in the
following way. We describe the radial velocity dispersion $\sigmaRi$,
as
\begin{equation}
  \sigmaRi^2(R)=\sigma_{0,i}^2\exp\left(-\sqrt{R^2+2\as^2}/\Rh\right),
\end{equation} 
(\citealt{Hernquist1993}), which implies that, far enough from the
center, $\sigmaRi^2 \propto \exp(-R/\Rh)$.  In the center the
smoothing parameter $\as$ reduces the velocity dispersion.
\cite{Hernquist1993} uses $\as=\Rh/4$. We prefer $\as=\Rh$, which
makes the smoothing much more gradual. We relate the tangential
velocity dispersion $\sigmaphii$ to the radial $\sigmaRi$ through the
epicyclic approximation, i.e.,
\begin{equation}\label{eq:epicycle}
  \sigmaphii^2(R)=\sigmaRi^2(R)\frac{\kappa^2(R)}{4\Omega^2(R)},
\end{equation} 
where $\Omega(R)$ and $\kappa(R)$ are the circular and epicyclic
frequencies (\citealt{BT2008}).

Finally, the vertical velocity dispersion $\sigmazi$ is obtained by
solving the following Jeans equation for an axisymmetric system
\begin{equation}\label{eq:sigmaz} \frac{1}{R}\frac{\partial\left[R
\rhodi(\overline{v_Rv_z})_i\right]}{\partial R}+
\frac{\partial\left(\rhodi\sigmazi^2\right)}{\partial
z}=\rhodi\frac{\partial\Phi}{\partial z},
\end{equation} where $\Phi(R,z)$ is the total potential of the Galaxy.
Assuming a velocity ellipsoid aligned with the $R$ and $z$ axes, i.e.,
$(\overline{v_Rv_z})_i=0$, the first term in the left-hand side of
Eq. (\ref{eq:sigmaz}) vanishes, and
\begin{equation}\label{eq:Jeansz} \sigmazi(R,z)=\frac{1}{\rhodi(R,z)}
\int_{z}^{\infty}\de z' \rhodi(R,z')\frac{\partial \Phi}{\partial z'}.
\end{equation} In practice, we compute Eq. (\ref{eq:Jeansz}) on a
discrete grid of equispaced points on the meridional plane and, to
obtain $\sigmazi(R,z)$ in a generic $(R,z)$ point, we linearly
interpolate between the nearest grid points.

Each velocity component of a particle is assigned by drawing a random
number from a Gaussian distribution with the respective dispersion.
The tangential component of the velocity, $v_{\phi,i}$, is corrected
for the asymmetric drift $\vai$,
\begin{equation}\label{eq:va} \vai(R)=\frac{\sigmaRi^2}{2\vc}
\left[\frac{\sigmaphii^2}{\sigmaRi^2}-1-\frac{\partial\ln(\rhodi
\sigmaRi^2)}{\partial \ln R}\bigg|_{z=0}\right],
\end{equation} (\citealt{BT2008}) where, again, we set
$(\overline{v_Rv_z})_i=0$.

The initial conditions generated in this way are not fully consistent
with the potential.  In particular we have assumed the velocity
distribution to be locally Gaussian. Therefore we expect these to
change in time, while reaching equilibrium with the potential.  These
transient effects are described in \cite{Minchev2009} and produce
kinematic arcs in the Solar Neighbourhood.  To avoid confusion between
these effects and those induced by the bar we first let our initial
conditions evolve in the axisymmetric potential for a time
$t_{\mathrm{r}}$.  We find that $t_{\mathrm{r}}\sim 8\Gyr$ suffices to
reach a stationary state around and inside the Solar radius.  We use
these evolved initial conditions as effective initial conditions for
our simulations with the bar.
\begin{figure}
  \centering
  \includegraphics[width=\columnwidth]{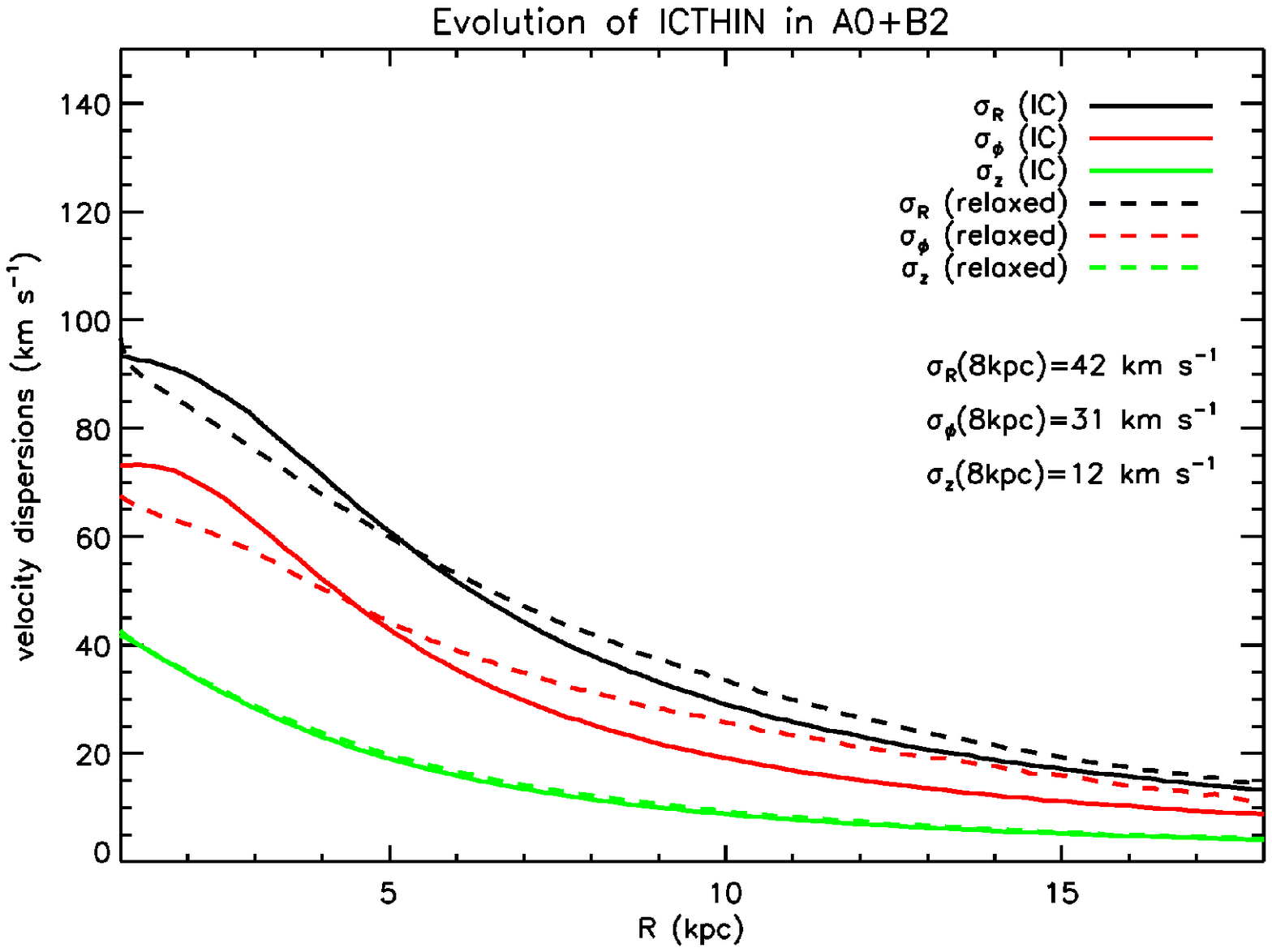}
  \centering
  \includegraphics[width=\columnwidth]{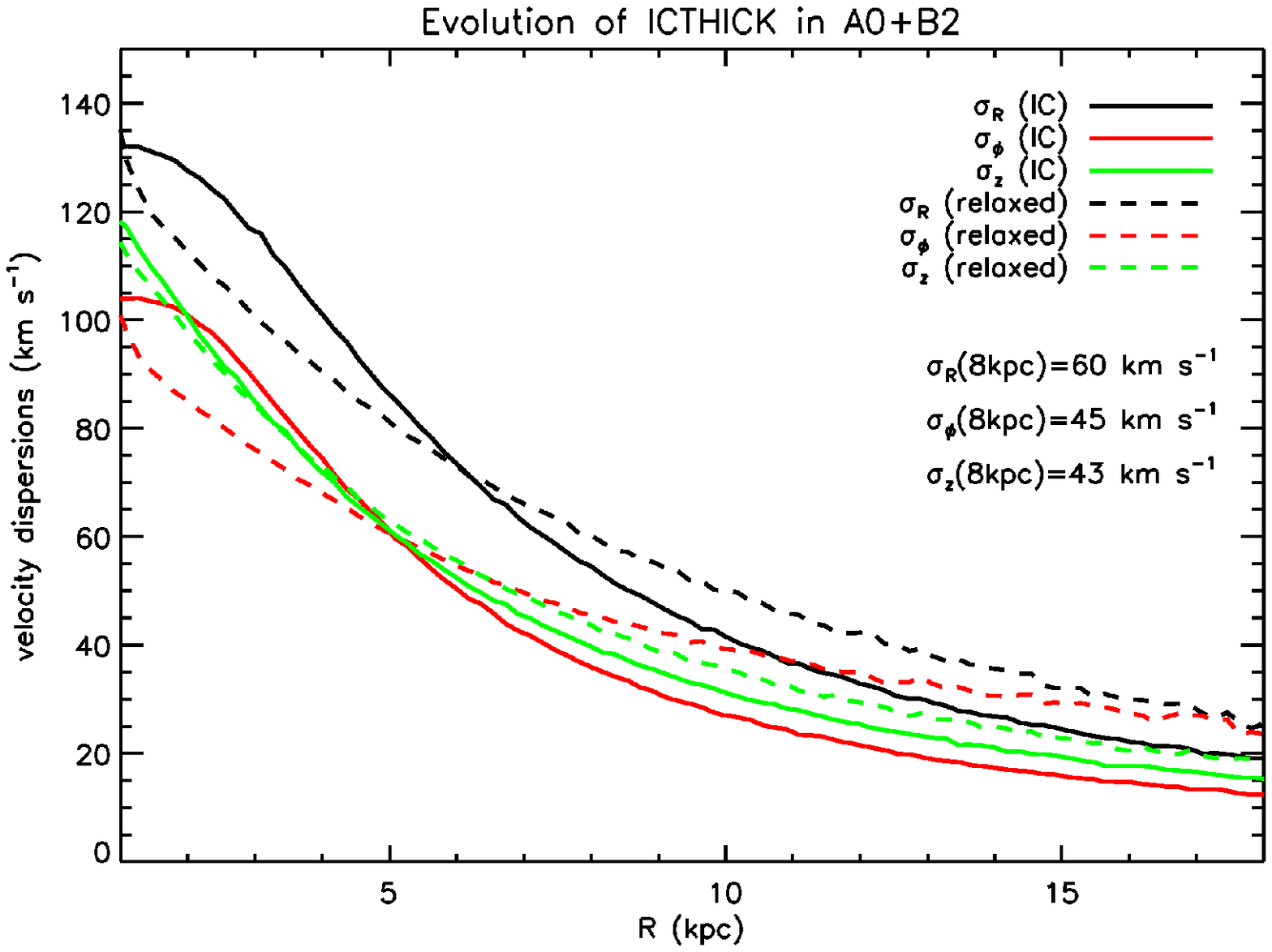}
  \caption{Velocity dispersions before and after the evolution
           of the initial conditions in the A0+B2 axisymmetric potential for
           $t_{\mathrm{r}}\simeq8.3\Gyr$ and for stars with 
           final $|z|<0.3\Kpc$. The final velocity dispersions at $R=8\Kpc$
           are also indicated.}\label{fig:sigmav}
\end{figure}

The evolution of the velocity dispersions for ICTHIN and ICTHICK is
shown in Fig. \ref{fig:sigmav} for the axisymmetric potential A0+B2
(the results for A0+B1 are similar).  We see that the radial and
tangential velocity dispersions increase in the outer parts of the
Galaxy and decrease in the inner regions, both for ICTHIN and ICTHICK.
In contrast, the vertical velocity dispersion does not evolve for
ICTHIN and only slightly increases for ICTHICK.  This is likely
because the vertical velocity dispersion is obtained directly by
solving the Jeans equation, while the radial dispersion and tangential
profiles are only imposed later using Eq.~(\ref{eq:epicycle}) and
related through the epicyclic approximation .

An intermediate age population in the thin disk ($\sim 5 \Gyr$) has
local velocity dispersions
$(\sigma_R,\sigma_\phi,\sigma_z)\sim(40,26,17)\kmsec$
\citep{Besancon,Holmberg2007}.  After evolution, our set of initial
conditions ICTHIN has comparable characteristics with slightly larger
$\sigma_\phi$ and smaller $\sigma_z$.  The measured local dispersions
of the stars in the thick disc are
$(\sigma_R,\sigma_\phi,\sigma_z)\sim(67,51,42)\kmsec$ according to
\cite{Besancon} or smaller in the $\phi$ and $z$ directions
$(\sigma_R,\sigma_\phi,\sigma_z)\sim(67,38,35)\kmsec$ as reported in
\cite{Bensby2003}.  Our initial conditions ICTHICK after the evolution
in the imposed potential are consistent with these measurements.

\subsection{Orbits and resonances}
\begin{figure}
  \centering
  \includegraphics[width=\columnwidth]{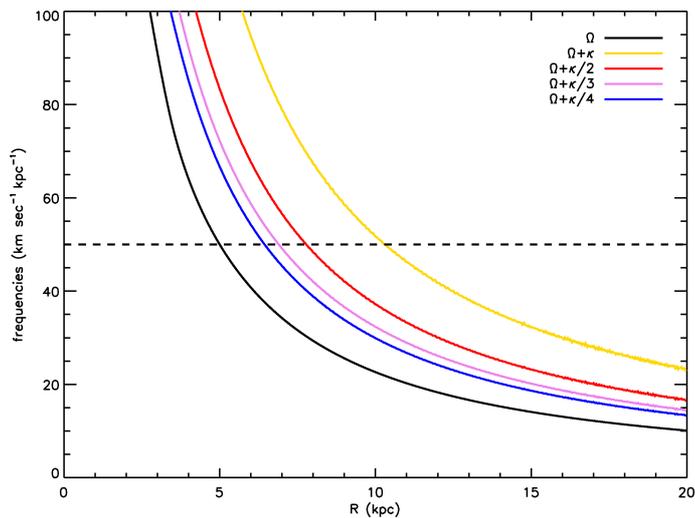}
  \caption{Positions of the main resonances for the model A0+GB2
    ($m=0$ Fourier component) with $\Omegab=50\kmseckpc$ (horizontal
    dashed line).  The position is given by the $R$ at which the
    curves cross the horizontal line.}\label{fig:om}
\end{figure}

Let us, for a moment, consider orbits in an axisymmetric potential.
In the epicyclic approximation (that is a good description only
  of orbits that are almost circular and with small vertical motion
  amplitudes) the azimuthal velocity of a star, located at $R$ in the
Galaxy, is
\begin{equation}\label{eq:epicyclic}
 %v_\phi\approx R\Omega(\Rg)\left[1-2\left(\frac{R}{\Rg}-1\right)\right],
  v_\phi\approx R\Omega(\Rg)\left(3-2\frac{R}{\Rg}\right),
\end{equation}
where 
$\Rg$ is the radius of the guiding center of its orbit.  When higher
order terms are considered, $v_\phi$ depends also on the value of
$v_R$ (e.g., \citealt{Dehnen1999}).  Of particular interest are the
orbits with $\Rg$ such that
\begin{equation}\label{eq:resonance}
  l_2\kappa(\Rg)=l_1\left[\Omega(\Rg)-\Omegab\right],
\end{equation}
where
$l_1$ and $l_2$ are two integer numbers.  For these orbits the
epicyclic frequency resonates with the frequency of the circular orbit
at $\Rg$, in the reference frame rotating with pattern speed
$\Omegab$.
In other words, orbits well described by the epicyclic approximation
(i.e., not too eccentric) that satisfy this condition are closed in
this reference frame.
Important for the rest of the paper are, especially, the Outer
Lindblad Resonance (OLR, $\Rg=\ROLR$) where $l_1=-2$ and $l_2=1$, the
$-1:1$ Resonance ($\Rg=\Roto$), where $l_1=-1$ and $l_2=1$, and the
Corotation Resonance ($\Rg=\Rcor$), where $l_2=0$ (i.e.,
$\Omega(\Rg)=\Omegab$).  In Fig. \ref{fig:om} we show the positions of
the main resonances for the A0 +GB2 potential but considering only the
$m=0$ Fourier component of GB2, for $\Omegab=50\kmseckpc$ (Table
\ref{tab:bar}).

Now, let us assume a small non-axisymmetric perturbation (the bar) to
the axisymmetric potential, rotating with pattern speed $\Omegab$.  If
the perturbation is small, the motion of most stars will still be
approximately described by the epicyclic theory. However, some stars
will be more strongly affected by the perturbation, namely those that
satisfy Eq. (\ref{eq:resonance}) in the axisymmetric potential.  As
stated by \cite{BT2008}, at the resonances the ``perturbation is
acting with one sign for a long time.  If the effects of a
perturbation can accumulate for long enough, they can become
important, even if the perturbation is weak''.

\subsection{Vertical motion} 

The simple dynamical picture that we described in the previous section
is complicated in two ways in the Milky Way: Galactic disks are 3D
structures, i.e. stars have vertical motions, and especially the thick
disk population is kinematically hot.

\begin{figure}
  \centering
  \includegraphics[width=\columnwidth]{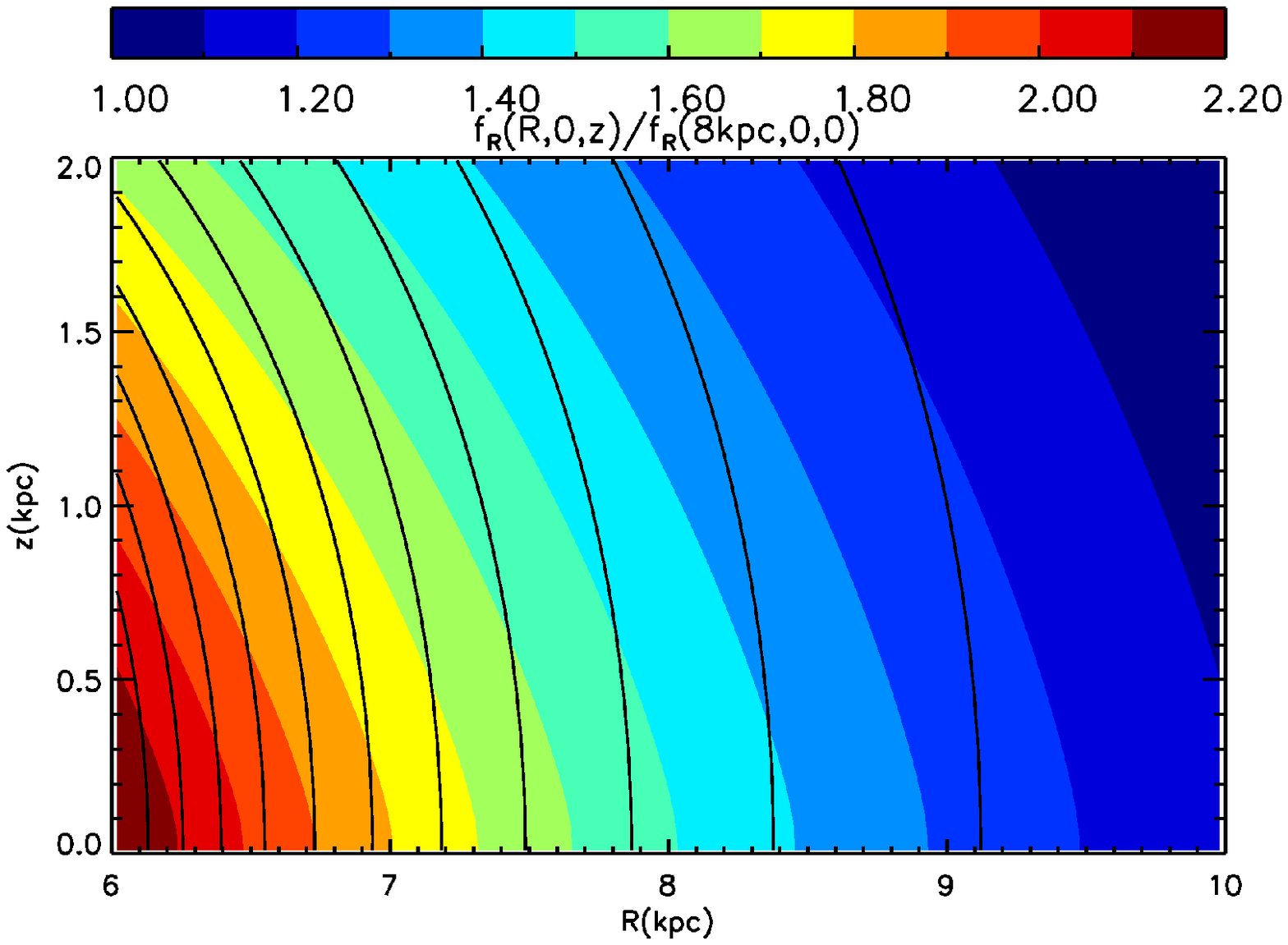}
  \includegraphics[width=\columnwidth]{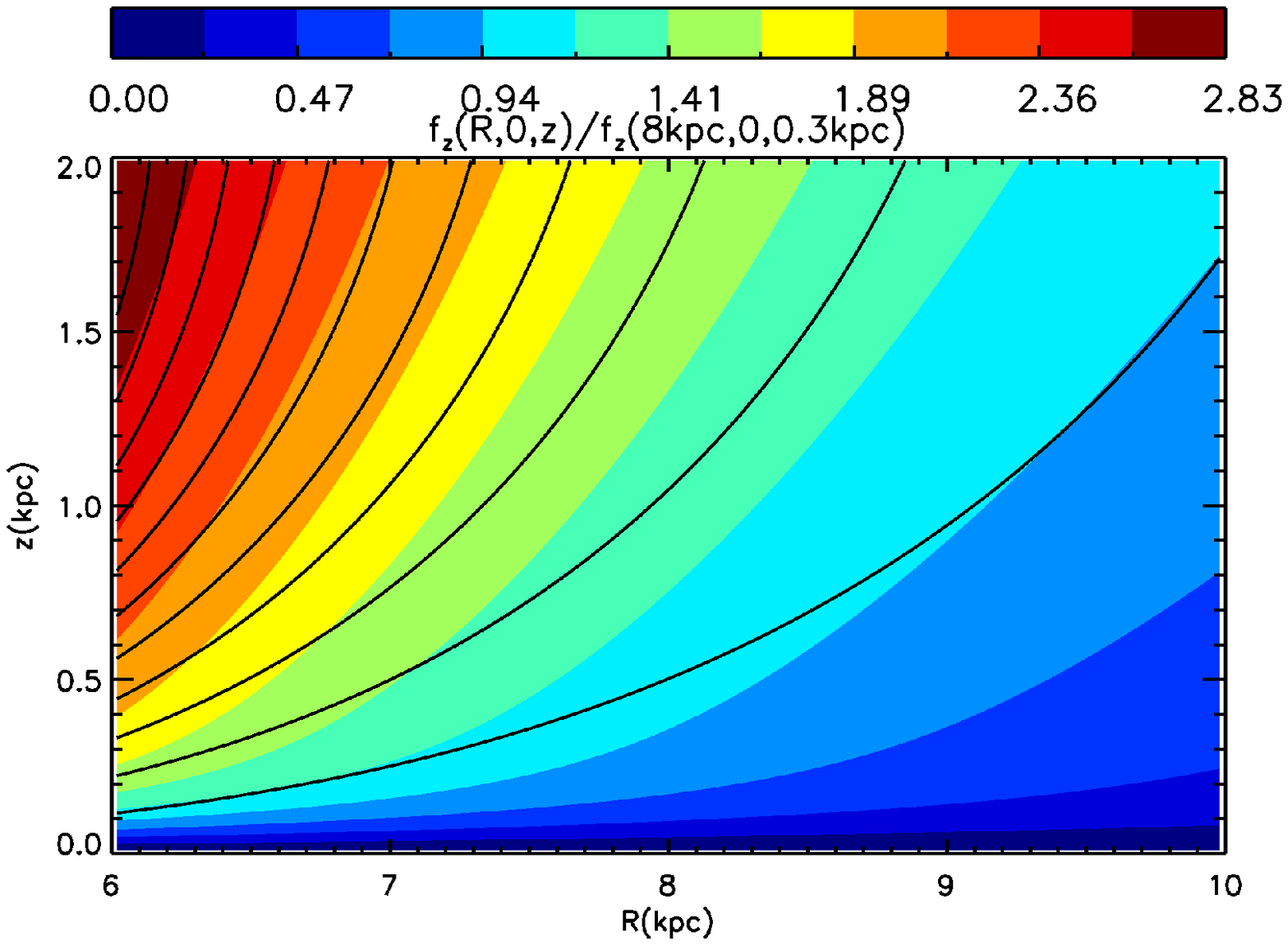}
  \caption{Isocontours of the $f_R$ (top panel) and $f_z$ (bottom
      panel) forces for the the potential A0+GB2 (colors) and $m>0$
      terms of the Fourier decomposition of GB2 (black solid lines).
      The meridional plane is shown at $\phi=0$ (along the long axis
      of the bar).}\label{fig:dec}
\end{figure}

The epicyclic approximation assumes that the orbits of the stars are
not too eccentric and that the amplitudes of the vertical motions are
so small that horizontal and vertical motions are nearly decoupled.
In an axisymmetric potential the motion is exactly decoupled when the
radial and vertical force ($f_R$ and $f_z$) are respectively functions
of $R$ and $z$ only. This is only true very near to the Galactic
plane.  In Fig. \ref{fig:dec} we have plotted the radial (top) and
vertical (bottom) forces for the total potential (colour) and the
contribution of the bar ($m>0$ terms of the Fourier decomposition,
black solid lines) for model A0+GB2. The bottom panel shows, as just
discussed, that $f_z$ is independent of $R$ only at small $z$ (where
it runs parallel to the cylindrical radius axis).  However, already
beyond $z \sim 0.3\Kpc$, it varies significantly with $R$, over the
range explored by many orbits that visit the Solar Neighbourhood. On
the other hand, the top panel shows that the $f_R$ isocontours are
less curved and that this force decreases more slowly with $z$ (and
this is even more true if we consider the bar's radial force only as
indicated by the black curves). For this reason the radial motion may
be expected to be more similar with height, than the vertical motion
with radius. Furthermore, it is important also to note that orbits
with angular momentum $L_z$ and vertical oscillations larger than
$\sim0.3\Kpc$ live in a range of $R$ that is hundred of parsecs
different from that of a planar orbit with the same $L_z$ angular
momentum (see \citealt{BinneyMcMillan2011}).

Also interesting is that for all our models the bar induces horizontal
forces that, at $R\sim8\Kpc$, are almost independent on $z$ (top panel
of Fig. \ref{fig:dec}, solid black curves). In fact, the relative
importance of the perturbation due to the bar slightly increases with
$z$, as a consequence of the decreasing radial force of the
axisymmetric part of the potential.  Therefore, we expect stars far
from the Galactic plane to experience a slightly stronger relative
effect of the bar compared to those on the plane.

The exact details of how the vertical motions, particularly for the
populations with larger kinematic temperature, affect the simple
dynamical picture previously sketched, require however, as especially
the bottom panel of Fig. \ref{fig:dec} shows, suitable and fully 3D
orbital integrations.

\section{Results} \label{sect:Results} Motivated by the above
discussion, we turn to numerical simulations to study how the bar
impacts the kinematics of stars near the Sun and neighbouring volumes.

We conventionally place the Sun on the Galactic plane, at
$(R,\phi)=(R_0,\phi_0)\equiv(8\Kpc,-20^{\circ})$ and we consider
cylindrical volumes throughout the Galaxy of radius $\rvol=0.3\Kpc$
and height $\hvol=0.6\Kpc$, centered at some $(R,\phi,z)$.  For each
volume we study the $v_R$ vs. $v_\phi$ velocity distribution.

\subsection{General effects of the bar and time evolution}
\begin{figure}
 \centering
 \includegraphics[width=\columnwidth]{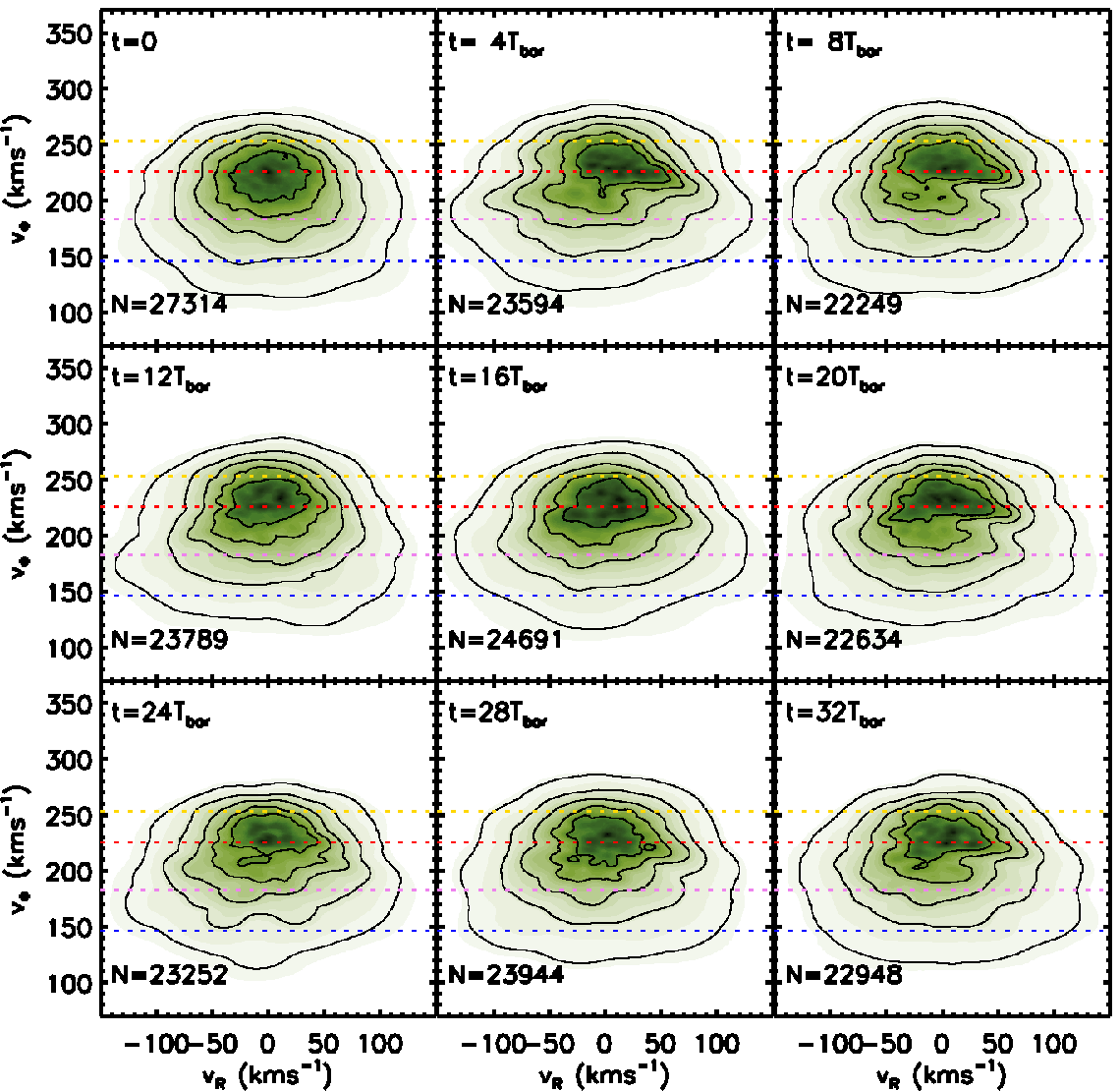}
 \caption{Evolution in time of the kinematics of the simulated LR thin
   disk in the Solar Neighbourhood volume for the potential
   A0+GB2. The dashed horizontal lines correspond to $\vphioto$
   (yellow), $\vphiOLR$ (red), $\vphitto$ (purple) and $\vphifto$
   (blue).  The density field is obtained with the modified Breiman
   estimator, described in \cite{Ferdosi2011}, with optimal pilot
   window $\sigma^\mathrm{opt}=9.6\kmsec$.  The contours enclose
   $0.25p_{1\sigma}\times i\ \%,\quad i=1,...,5$ of the
   probability, where $p_{1\sigma}=68.268$. In the lower left part of
   the panel we indicate the number of particles in the
   volume.}\label{fig:time_thin}
\end{figure}

In Fig. \ref{fig:time_thin} we show the evolution in time of the
kinematics in the Solar Neighbourhood volume for the model A0+GB2 and
the LR thin disk initial conditions (LR ICTHIN).
Specifically, to describe the kinematics we use a probability density
function of $(v_R,v_\phi)$, determined with the modified Breiman
  estimator, described in \cite{Ferdosi2011}, with optimal pilot
  window $\sigma^\mathrm{opt}=9.6\kmsec$.  We show $5$ density
contours that enclose (from inside to outside) $0.25p_{1\sigma}\times
i\ \%,\ i=1,...,5$ of the probability, where
$p_{1\sigma}=68.268$.  The $4$-th contour from inside out corresponds,
therefore, to the $1\sigma$ contour.  The evolution seen in
Fig. \ref{fig:time_thin} is qualitatively similar for all the other
potentials (and respective values of the parameters) explored.  As we
can see, the form of the distribution changes with time (see
\citealt{Minchev2010} for a colder disk).  Initially, the distribution
of stars is featureless, symmetric in $v_R$, and with a tail of stars
at low $v_\phi$ due to the asymmetric drift.  At $t=4\Tbar$ the
distribution has changed completely: it is split up in two regions and
presents other deformations that make it asymmetric, both in $v_R$ and
$v_\phi$.

Before proceeding with the analysis of the evolution in this plane, it
will be useful for the rest of the paper to introduce a terminology
describing the features seen in velocity space.  In
Fig. \ref{fig:time_thin} we have indicated the $v_\phi$ of four
important families of resonant orbits, using Eq. (\ref{eq:epicyclic})
for $R=8\Kpc$: $\vphioto$ (yellow dashed line), $\vphiOLR$ (red dashed
line), $\vphitto$ (purple dashed line) and $\vphifto$ (blue dashed
line).  As we can see, the %{\color{red} deep}
diagonal valley in the velocity distribution corresponds roughly to
$\vphiOLR$, confirming the large effect of the Outer Lindblad
Resonance on the kinematics of the stars.  Following
\cite{Dehnen2000}, we call the peak in the density of stars above the
valley the ``LSR Mode'', and the one below the ``OLR
Mode''. Furthermore, in the right part of the LSR Mode, at
$(v_R,v_\phi)\sim(75\kmsec,220\kmsec)$, there is an elongation in
$v_R$, which we call ``Horn''.

Previous studies have used default integration times of $t_2=4\Tbar$
(\citealt{Dehnen2000}). We notice here that between $t=4\Tbar$ and
$t=16\Tbar$ there is still significant evolution of the distribution,
which only reaches a stationary configuration later. For instance, the
valley between the LSR Mode and the OLR is progressively filled with
stars. The OLR and LSR Modes change shape in time.

In general, it is clear from Fig. \ref{fig:time_thin} that all the
features associated to the bar are most prominent for short
integration times.  The features that we find in the velocity
distribution for these times are %{\color{red} very}
similar to those observed by \cite{Dehnen2000} in 2D integrations with
a quadrupole bar.  This is the first time that they are seen in 3D
simulations.  On the other hand, the time evolution confirms, again in
3D, the results obtained by \cite{Fux2001} in his study of the
phase-mixing of orbits in 2D and with the quadrupole bar.

As our thin disk %simulations ($\sigma_R \sim 45 \kmsec$)
corresponds to an intermediate-old population of
stars, %($\sim 10 \Gyr$; see \citealt{Besancon}),
it is unlikely that they experienced the effects of the bar only for
$4\Tbar$ ($\approx 492\Myr$).  On the other hand,
\cite{ColeWeinberg2002}, using properties of infrared carbon stars
from the Two Micron All Sky survey, state that the Milky Way Bar is
likely to be younger than $3\Gyr$, fixing an upper limit for its
formation to $6\Gyr$ ago (however studies of the bulge indicate that
it is dominated by a stellar population older than $10\Gyr$, e.g.,
\citealt{MinnitiZoccali2008}).  Given this, we decide to choose a
default integration time of $t_2=24\Tbar\sim3\Gyr$, when the
  kinematics have roughly reached stationarity.
Note that we are effectively assuming that the properties of the bar,
such as its pattern speed did not change much during this time.

\subsection{Thin disk}

\begin{figure}
 \centering
 \includegraphics[width=\columnwidth]{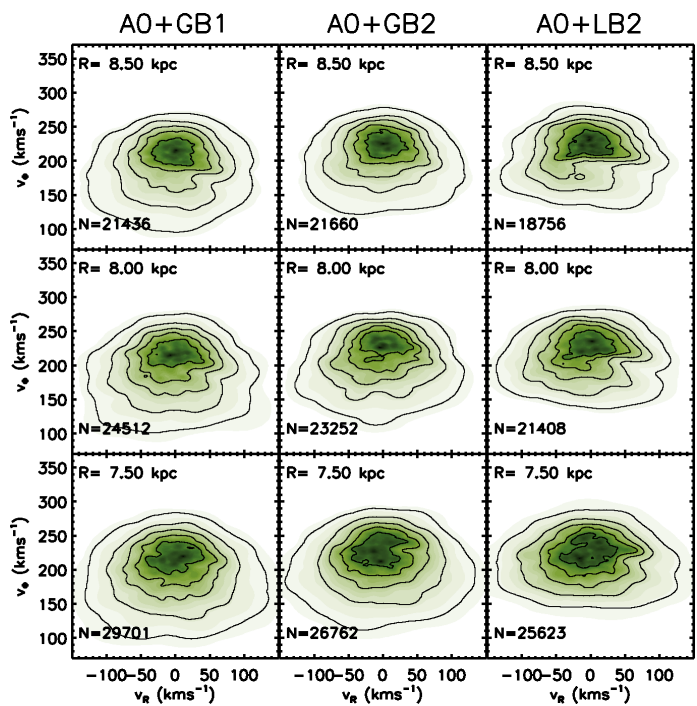}
 \caption{The effect of different bar potentials on the kinematics of
   the simulated thin disk in 3 volumes centered at an angle
   $\phi=-20^{\circ}$ and $R=8.5,8,7.5\Kpc$ for
   $t=24\Tbar$.}\label{fig:eff_thin}
\end{figure}

Fig. \ref{fig:eff_thin} shows the kinematics of the thin disk in the
Solar Neighbourhood and in two nearby volumes aligned with the Solar
Neighbourhood ($\phi=-20^{\circ}$) but at different radii ($R=8.5\Kpc$
and $R=7.5\Kpc$) for various potentials and LR simulations.  The first
and second column show the distributions with the bar potential with
$\Mbar=10^{10}\Msun$ and $\Mbar=2\times10^{10}\Msun$,
respectively. Not surprisingly, the effect of the least massive bar is
weaker compared to a bar twice as massive.  The density contrast
between the OLR Mode and the LSR Mode is smaller, and the Horn is
slightly less prominent for the least massive bar. The overall
distribution has also a more symmetric appearance. These small
differences are explainable by simply noting that the force of the
perturbation only slightly smaller in the case of the least massive
bar. The non-axisymmetric part of the force (i.e. excluding the
monopole term associated to the bar) differs only by $\sim 30\%$ at
$R=8\Kpc$ between the GB1 and GB2. Some of the differences between the
distributions may also be ascribed to the difference between the
circular velocity curves of A0+B1 and A0+B2, which have the resonances
at slightly different positions.  For instance, for
$\Omegab=50\kmseckpc$, for A0+GB1 and for A0+GB2 (see Table
\ref{tab:bar}).  Since the resonances are mostly responsible for the
effects of the bar on the kinematics of the stars, we expect some
differences at fixed $R$ \footnote{These differences may also reflect
  a dissimilar evolution of the distributions. As explained in
  \cite{Minchev2010}, the typical time of libration of the stars
  trapped to the resonances depends on the bar strength; since the
  shape of the distributions is defined mostly by the resonances, we
  may expect, at the same snapshot, different evolutionary stages for
  different bar strengths.}.

The second and third columns of Fig. \ref{fig:eff_thin} allow us to
compare more massive bar models ($\Mbar=2\times 10^{10} \Msun$) with
different geometrical properties (GB and LB).  The Long Bar (third
column) has slightly different effects on the thin disk velocity
distributions than the less elongated Galactic Bar. It produces
sharper features such as a Horn with a more defined lower edge,
especially for $R=8.5\Kpc$ and $R=8\Kpc$.
This was already noticed by \cite{GardnerFlynn2010} when they compared
the responses to these two Ferrers bar models, but we confirm here
their results in our 3D simulations.  Nevertheless, we stress that,
for this long integration time, the differences between the effect of
the different bars on the kinematics inside the various volumes
explored is much less significant than for shorter integration
times. Furthermore, for short integration times the structures are
much more evident in all our bar models.

Given the uncertainties in having a satisfactory model for the bar
(recent studies even suggest that it could be a superposition of
something similar to our Galactic Bar and Long Bar, see
\citealt{Robin2012}), we use from now on as a default model the
Galactic Bar and the potential A0+GB2, confident that the effects will
be similar for all the other models, especially for long integration
times, as illustrated in Fig. \ref{fig:eff_thin}.

\begin{figure}
  \centering
  \includegraphics[width=\columnwidth]{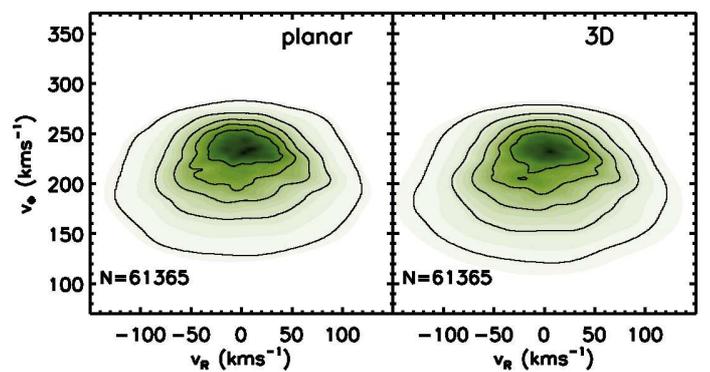}
  \caption{Kinematics in the Solar Neighbourhood at $t=24\Tbar$ for
    two subsets of the thin disk HR simulation with A0+GB2
    potential. Left panel: the most planar particles in the
    simulation.  Right panel: a random subsample of the simulation
    with the same number of stars.}\label{fig:plane}
\end{figure}
Since most works so far have used 2D simulations, it is important to
understand if significant differences exist for the kinematics near
the plane in the 2D vs 3D simulations.  To this end we consider a
subset of the HR 3D thin disk simulation in the potential
A0+GB2, constituted by those particles that, at $t=0$, have
$|z|<0.05\Kpc$ and $|v_z|<5\kmsec$. These are the particles whose
vertical oscillations have the smallest amplitude in the simulation:
at $t=24\Tbar$ their $z$ dispersion is still less than $0.04\Kpc$ and
their $v_z$ dispersion less than $4\kmsec$.  It is then reasonable to
expect they behave in a similar fashion to particles in a 2D
simulation, distributed on the Galactic plane with similar $(R,\phi)$
density and kinematics at $t=0$, and subject to the same potential
A0+GB2.

Fig. \ref{fig:plane} shows the kinematics of the subset of most
``planar'' particles (left panel) and the kinematics of a random
subsample of the whole 3D simulation with the same number of stars
(right panel), in the Solar Neighbourhood volume at the default
integration time.  We note that there are no large differences between
the velocity distributions for these two data sets. The two sets do
have slightly differing velocity dispersions: the planar particles are
slightly colder ($\sigma_R=42\kmseckpc$, while $\sigma_R=44\kmseckpc$
for the 3D subsample), but this is a reflection of the initial
conditions.
This agreement stems likely from the fact that the planar and vertical
motions are decoupled for most thin disk particles.  Since our 3D thin
disk is particularly cold in $v_z$ ($\sigma_z \sim 12 \kmsec$ at
$R=8\Kpc$), it is also possible that this is more true in our
simulations than in reality since the observed local $\sigma_z \sim 17
\kmsec$ \citep{Besancon,Holmberg2007}.
\begin{figure*}
  \centering
  \includegraphics[width=0.8\textwidth]{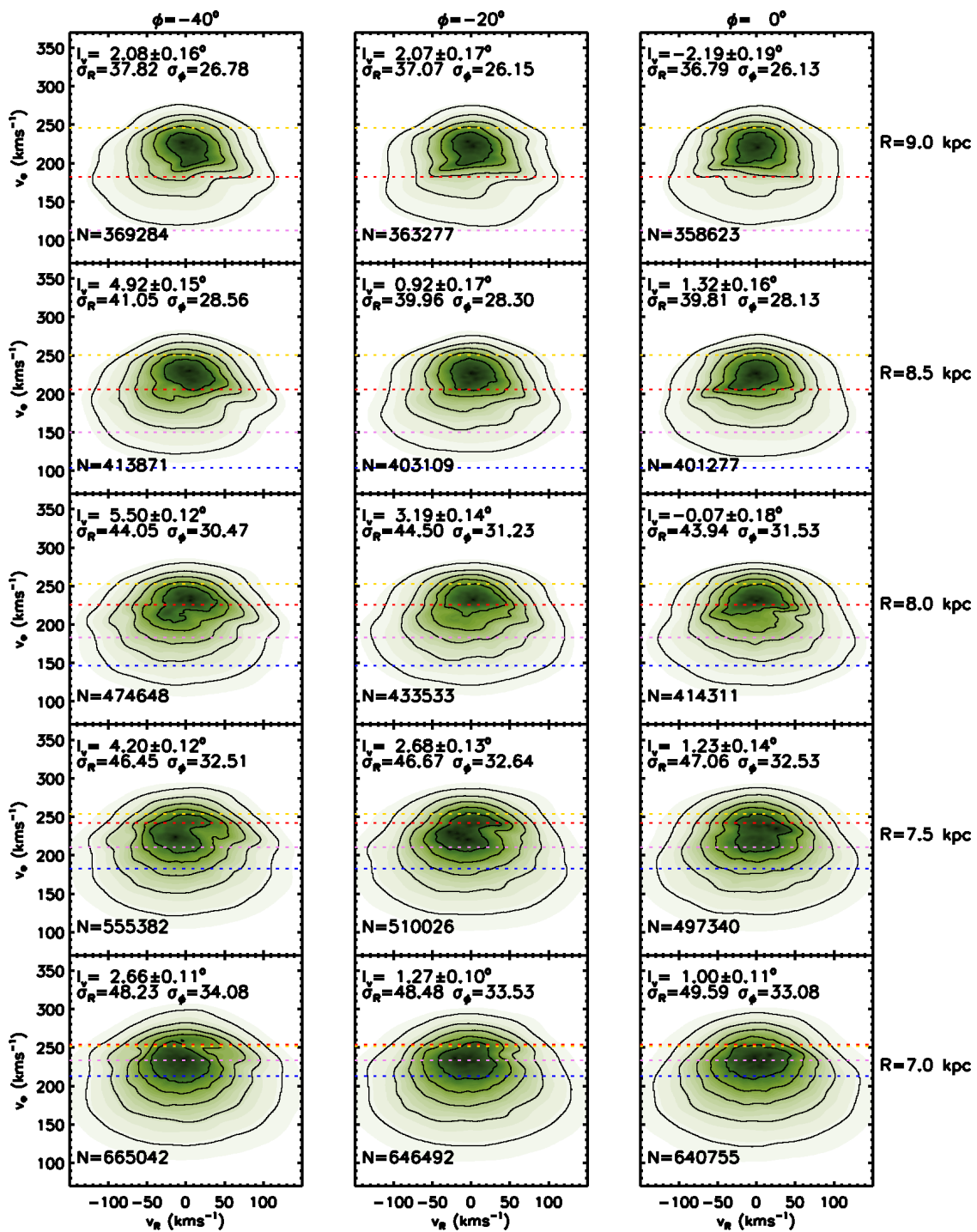}
  \caption{Kinematics of the simulated HR thin disk in the
    potential A0+GB2 in different volumes at $t=24\Tbar$. Different
    columns correspond to volumes centered at different angles from the
    bar in the direction of antirotation, from $\phi=-40^{\circ}$
    (leftmost column) and with step $\Delta\phi=20^{\circ}$.
    Different rows correspond to volumes centered at different
    Galactocentric radii, from $R=9\Kpc$ (topmost row) and with step
    $\Delta R=0.5\Kpc$.  The dashed lines correspond to $\vphioto$
    (yellow), $\vphiOLR$ (red), $\vphitto$ (purple) and $\vphifto$
    (blue). Notice how for the volumes in the bottom row $\vphiOLR$
    and $\vphioto$ are superposed.} \label{fig:thin_panel}
\end{figure*} 

In Fig. \ref{fig:thin_panel} we show the kinematics of the thin disk
in the vicinity of the Solar Neighbourhood at $t=24\Tbar$ and in HR.
From left to right, the columns correspond to volumes centered at
angles $\phi=-40^{\circ}$, $\phi=-20^{\circ}$ and $\phi=0$ (aligned
with the bar).  From bottom to top, the rows correspond to different
Galactocentric radii from $R=7\Kpc$ to $R=9\Kpc$ in steps of $\Delta
R=0.5\Kpc$.  The yellow, red, purple and blue dashed lines represent
$v_\phi$ of the main resonances (as in Fig. \ref{fig:time_thin}),
obtained by using Eq. (\ref{eq:epicyclic}) with the $R$ of the
corresponding volume.  This figure shows that the main kinematic
features associated to the bar vary as a function of position in the
Galaxy, similarly to the effects already reported in the literature
for 2D simulations.  First, we note that the main features shift in
$v_\phi$ as $R$ varies.  At $R=9\Kpc$ (first row), $\vphiOLR$ is in
the low $v_\phi$ tail of the distribution ($v_\phi\sim180\kmsec$), the
OLR Mode has a small extension, and the Horn is placed at the lower
right edge of the $1\sigma$ contour ($4$-th contour).
For $R\sim 7.5$ and $8\Kpc$ $\vphiOLR$, and in consequence the
separation between LSR Mode and OLR Mode, lies at the center of the
$v_\phi$ distribution ($v_\phi\sim220\kmsec$), and the LSR Mode and
OLR Mode become clearly separated.  For $R=7\Kpc$ (bottom
row), %$\vphiOLR$ and $\vphioto$ are superposed,
the LSR mode can hardly be seen %becomes very small,
and the Horn is now at the upper right edge of the distribution.
Since these features are mostly associated with the resonant orbits,
their shift can be explained in light of Eq. (\ref{eq:epicyclic}), for
each of the resonant families ($\Rg$ constant).
\begin{figure}
  \centering
  \includegraphics[width=\columnwidth]{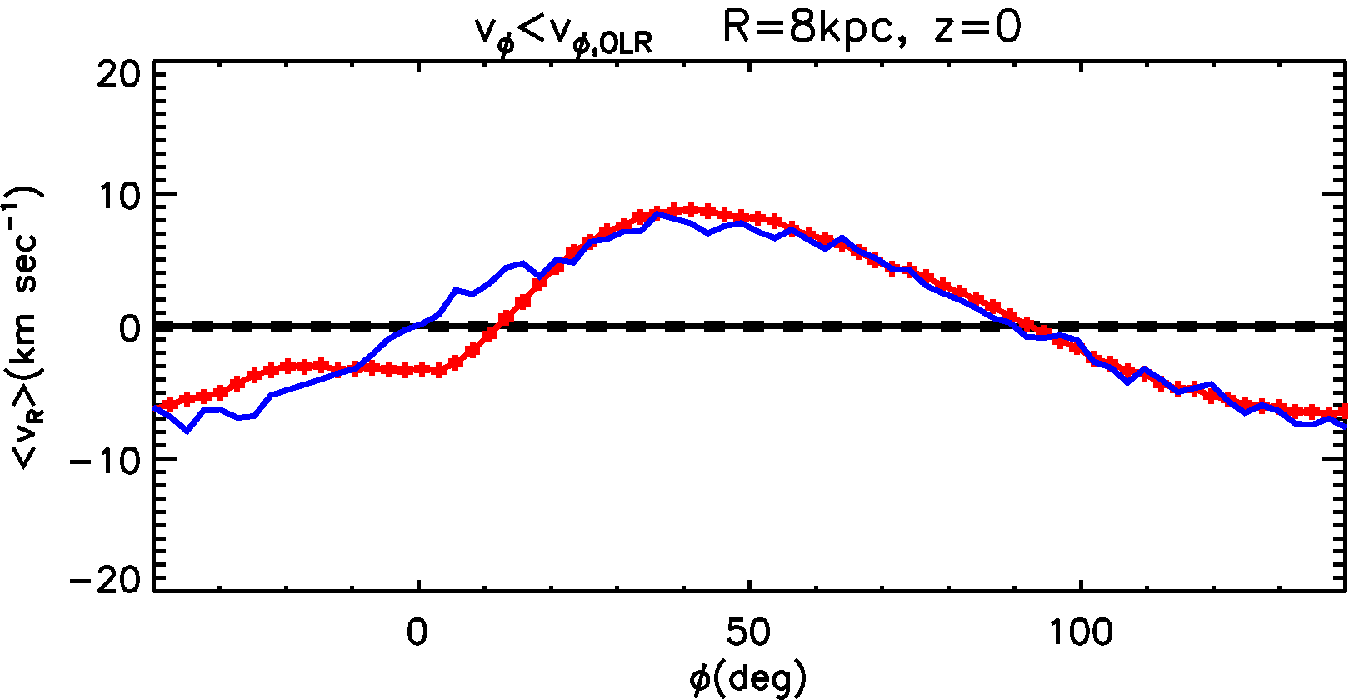}
  \includegraphics[width=\columnwidth]{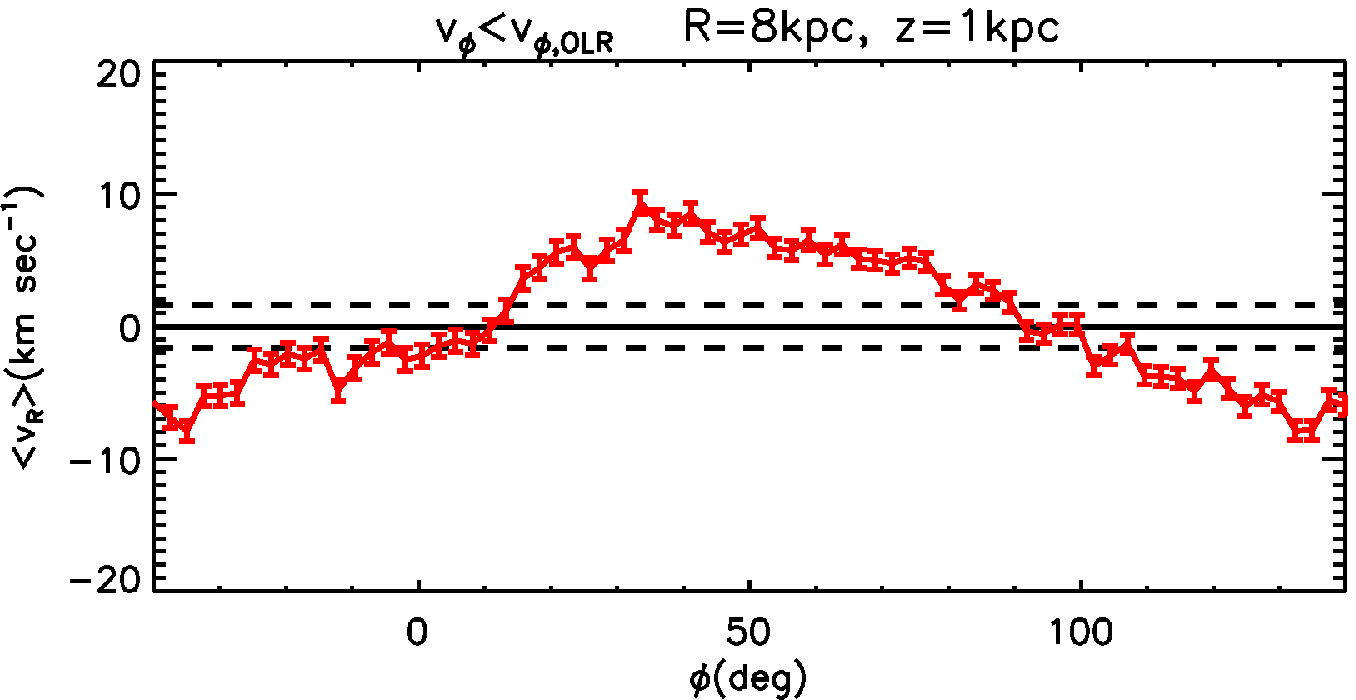}
  \caption{Average $v_R$ of thin disk particles $v_\phi<\vphiOLR$ in
    volumes centered at $R=8\Kpc$ as a function of $\phi$ for the HR
    simulation A0+GB2 , at $t=24\Tbar$ (solid red).  For comparison we
    show with the solid black line the average $v_R$ over all angles
    $\phi$ for the axisymmetric initial conditions ICTHIN, and with
    dashed lines twice the standard deviation around this average. The
    top and bottom panels are for $z=0\Kpc$ and $z=1\Kpc$
    respectively.  The blue line in the top panel represents the
    average $v_R$ of thin disk particles $v_\phi<\vphiOLR$ in volumes
    centered at $R=8\Kpc$ as a function of $\phi$ for the LR
    simulation A0+GB2, at $t=120\Tbar$ (solid blue).
  }\label{fig:neg_vR}
\end{figure}

\begin{figure}
  \centering
  \includegraphics[width=\columnwidth]{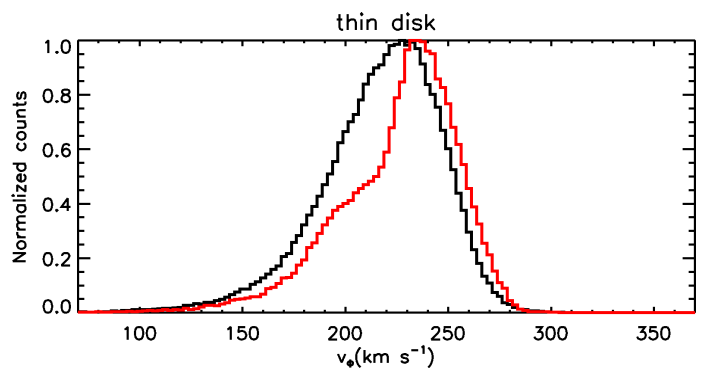}
  \caption{The distribution of thin disk stars in $v_\phi$ for the
    volume centered at $(R,\phi,z)=(8\Kpc,0,0)$ and for stars with
      $\left|v_R\right|<30\kmsec$. The red line represents the
    simulation A0+GB2, at $t=24\Tbar$. The solid black line represents
    the axisymmetric initial conditions.}\label{fig:hist_thin}
\end{figure}

We can also observe a change in the mean the $v_R$ of the OLR Mode as
the orientation of the volume with respect to the bar ($\phi$) is
varied.  While the OLR Mode is almost completely symmetric with
respect to the $v_R$ axis for $\phi=0$, it moves towards $v_R<0$ for
more negative $\phi$. At $\phi=-40^{\circ}$ the OLR Mode is completely
in the $v_R<0$ half plane at almost all radii.  We quantify this
change in Fig. \ref{fig:neg_vR}.  The red solid line shows the mean
$v_R$ for particles with $v_\phi <
\vphiOLR$, % (most of the composition of the OLR mode),
in volumes at $R=8\Kpc$ as function of $\phi$ (plotted in steps
$\Delta\phi=2.5^{\circ}$).  The black solid line represents the same
quantity, but for the axisymmetric initial conditions and averaged
between all the angles. The dashed lines represents twice the standard
deviation of the set of means obtained for all the angles.  This
figure shows that the OLR Mode varies periodically in $v_R$ as a
function of $\phi$, and that the amplitude of this oscillation is
significantly above the noise level. Notice however how the curve is
not perfectly symmetric: there is a wiggle for $-40^{\circ}<\phi<0$,
and the amplitude is different between positive and negative $\langle
v_R\rangle$. Because of the symmetry of the potential, we would expect
instead $\langle v_R \rangle(\phi)=-\langle v_R \rangle(\pi-\phi)$,
for each $\phi$. This happens because the disk has not reached a fully
stationary equilibrium configuration at $R=8\Kpc$ at this time
(\citealt{Fux2001,Muhlbauer2003}). A much longer (and unrealistic)
integration time of $120\Tbar$ ($\sim 15\Gyr$) is required to reach
complete symmetry (Fig. \ref{fig:neg_vR}, blue line).  For volumes
approximately aligned or perpendicular to the bar the OLR Mode has a
null mean, that is the OLR Mode is centered.  For these cases,
however, the behaviour is still significantly different from the
axisymmetric case, as one can see directly in the right column of
Fig. \ref{fig:thin_panel}, at $R=8\Kpc$, where the bimodality, despite
being centered with respect to $v_R$, is still present, and as shown
in Fig. \ref{fig:hist_thin}. In this figure we restrict the attention
to those stars with $\left|v_R\right|<30\kmsec$, where the differences
from the axisymmetric initial conditions are enhanced.
\begin{figure}
  \centering
  \includegraphics[width=\columnwidth]{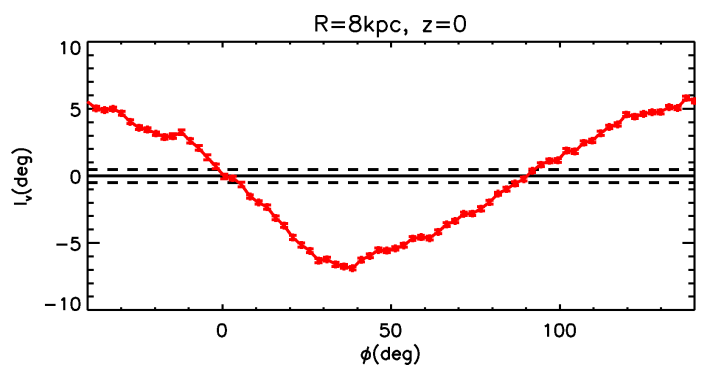}
  \caption{Vertex deviation in volumes centered at $R=8\Kpc$ as a
    function of $\phi$.  The description of the lines is the same of
    Fig. \ref{fig:neg_vR}.}\label{fig:lv}
\end{figure}

The vertex deviation $l_v$ quantifies the orientation of the velocity
ellipsoid on the $z=0$ plane. Following \cite{BinneyMerrifield} this
is:
\begin{equation}
  \lv\equiv\frac{1}{2}\arctan 
  \left[\frac{2\langle v_R \left(v_\phi-\langle v_\phi\rangle\right)\rangle}
    {\sigma_R^2-\sigma_\phi^2}\right].
\end{equation}
The corresponding values are quoted in all panels of
Fig. \ref{fig:thin_panel}.  In the Solar Neighbourhood vicinity $\lv$
is generally small. This means that the velocity ellipsoid is only
slightly misaligned with the respect to the $v_R$ and $v_\phi$ axis.
The vertex deviation behaviour is fully quantified in
Fig. \ref{fig:lv}.  For $R=8\Kpc$, we see that $\lv$ is positive for
$\phi<0$, null for $\phi=0$ (volumes aligned with the bar), and
negative for $\phi>0$, until it changes again sign at
$\phi=90^{\circ}$.  This is consistent with the results of
\cite{Muhlbauer2003}, who showed that the bar can cause significant
periodic perturbations in the velocity moments throughout the Galaxy
for the 2D case.
\begin{figure}
  \centering
  \includegraphics[width=\columnwidth]{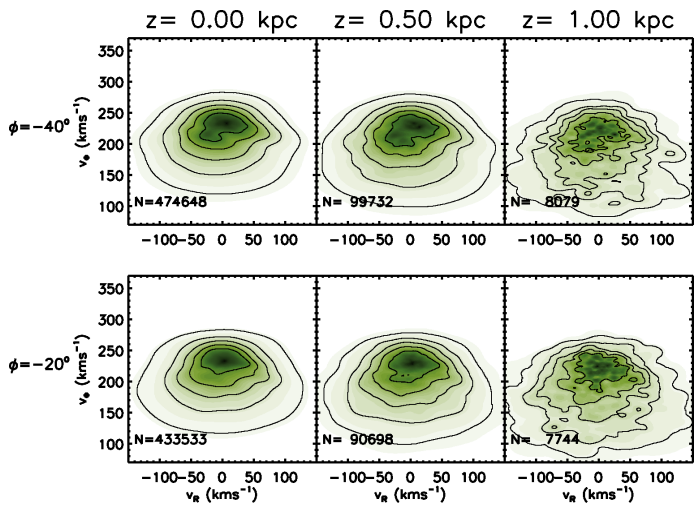}
  \caption{Kinematics of the thin disk HR simulation in the
    potential A0+GB2 in volumes centered at $R=8\Kpc$ and
    $\phi=-20^{\circ}$ and $\phi=-40^{\circ}$ and $z=0,0.5,1\Kpc$
    (first, second and third column
    respectively).} \label{fig:evz_thin}
\end{figure}

Our 3D simulations offer us the unique opportunity to trace the disk
kinematics far from the Galactic plane.
In Fig. \ref{fig:evz_thin} we show the velocity distribution of the
thin disk for the default bar model, for volumes at $R=8\Kpc$, at
angles $\phi=-20^{\circ}$ (top row) and $\phi=-40^{\circ}$ (bottom
row), but centered at different $z$.  For both angles it is possible to
trace the characteristic features (LSR and OLR Mode, Horn) present at
$z=0$, at least up to $z=0.5\Kpc$. %Although the dispersion increases,
This is confirmed by the periodic oscillation of the average $v_R$ of
the OLR Mode which is significant, even at this height.  Periodicity
is present also at $z=1\Kpc$ (see lower panel of
Fig. \ref{fig:neg_vR}).

We mentioned in earlier sections how the relative strength of the bar
slightly increases with $z$. It might be then not surprising to see
effects of the bar at these heights. However the volumes at these
heights are populated very differently than the ones located on the
plane: the orbits are more eccentric, they have larger vertical
amplitudes and they rotate more slowly in average ($\langle
v_\phi\rangle(8\Kpc,20^{\circ},0)-\langle
v_\phi\rangle(8\Kpc,20^{\circ},1\Kpc) \sim 20\kmsec$). It seems,
however, that the differences in the velocity distributions at various
heights are not much driven by the different types of orbits but
mostly by the much smaller numbers of stars (and related Poisson
noise) far from the plane.

\subsection{Thick disk}
\begin{figure*}
  \centering
  \includegraphics[width=0.8\textwidth]{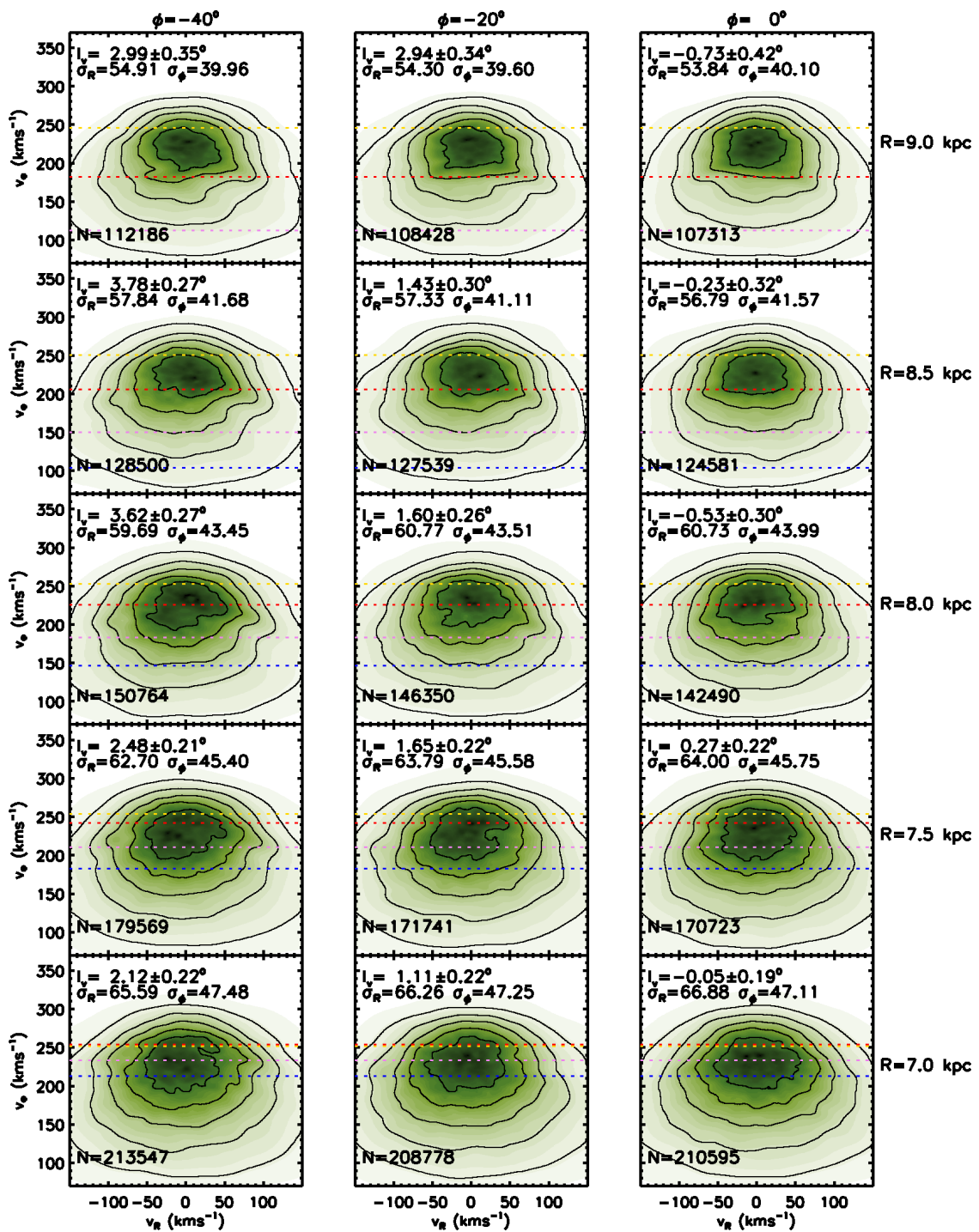}
  \caption{Kinematics of the simulated thick disk in the potential
    A0+GB2 in different volumes at $t=24\Tbar$. As in
    Fig.~\ref{fig:thin_panel} the columns correspond to volumes
    centered at different angles, and the rows to volumes centered at
    different Galactocentric radii.  The dashed lines indicate to
    $\vphioto$ (yellow), $\vphiOLR$ (red), $\vphitto$ (purple) and
    $\vphifto$ (blue).}\label{fig:thick_panel}
\end{figure*} 
Even in a hot and vertically extended population like the thick disk,
it is still possible to recognize the dynamical imprints of the bar on
its kinematics.  Fig. \ref{fig:thick_panel} (equivalent for the thick
disk to Fig. \ref{fig:thin_panel}) shows that, for
$\phi\leq20^{\circ}$, we can find the Horn in all volumes (see the
$1\sigma$ contour) at larger $v_\phi$ for decreasing $R$.  At the same
time, we can recognize the OLR Mode as a distinct feature with the
respect to the LSR mode in several volumes. Specifically, for
$R=8\Kpc$ and $R=8.5\Kpc$ the fourth contour shows a lack of stars at
positive $v_R$ in the range $150\kmsec<v_\phi<220\kmsec$.

\begin{figure}
  \centering
  \includegraphics[width=\columnwidth]{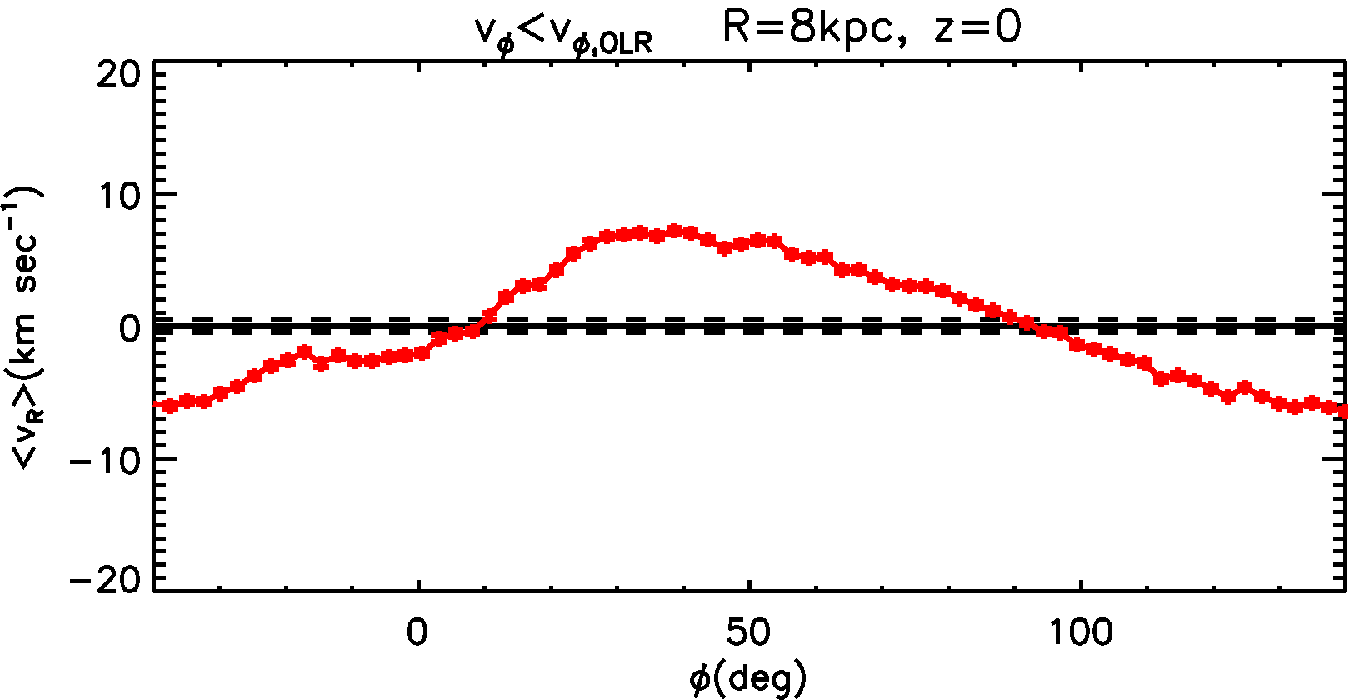}
  \includegraphics[width=\columnwidth]{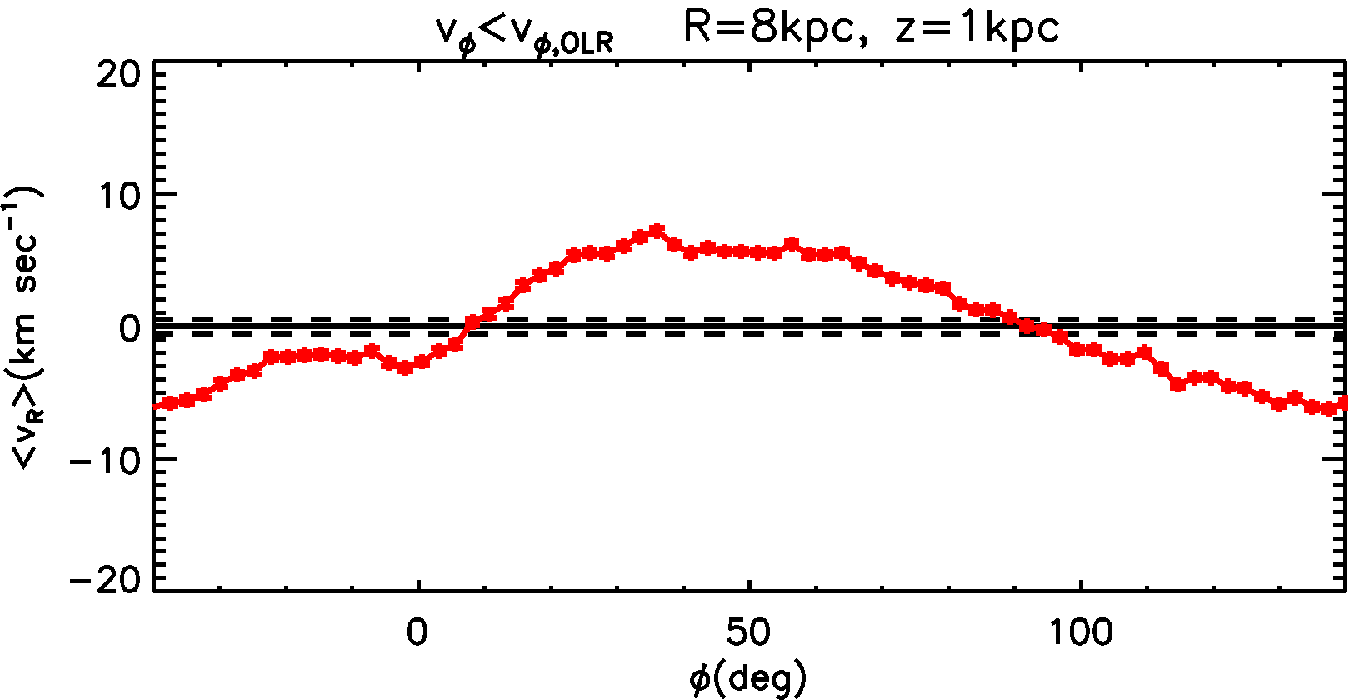}
  \includegraphics[width=\columnwidth]{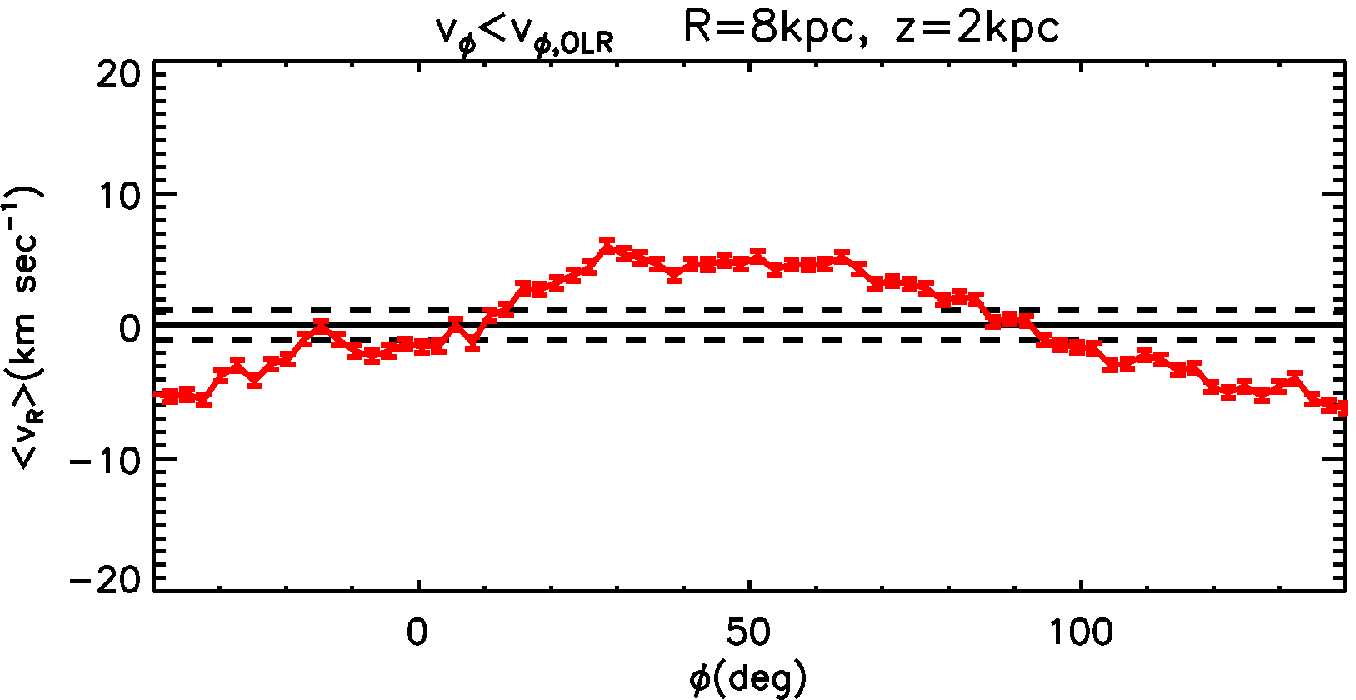}
  \caption{Average $v_R$ of thick disk particles $v_\phi<\vphiOLR$ in
    volumes centered at $R=8\Kpc$ as a function of $\phi$, for
    simulation A0+GB2, at $t=24\Tbar$ (solid red).  As in
    Fig.~\ref{fig:neg_vR} we show for comparison the average (solid
    black) and twice the standard deviation (dashed black) $v_R$ for
    the axisymmetric initial conditions ICTHICK.  The top, middle and
    bottom panels correspond to $z=0\Kpc$, $z=1\Kpc$ and $z=2\Kpc$
    respectively.}\label{fig:neg_vR_thick}
\end{figure}

\begin{figure}
  \centering    
  \includegraphics[width=\columnwidth]{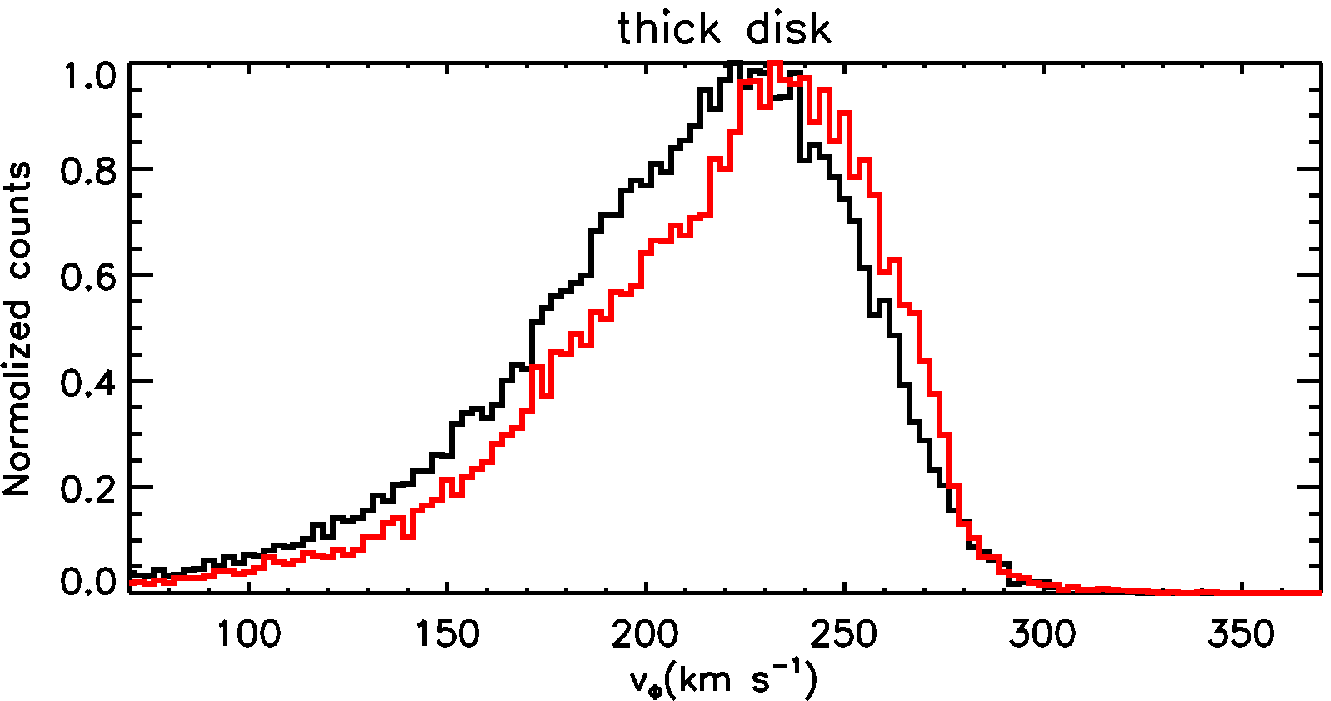}
  \caption{The $v_\phi$-distribution of thick disk stars for the
    volume centered at $(R,\phi,z)=(8\Kpc,0,0)$ and for stars with
      $\left|v_R\right|<30\kmsec$. The red line represents the
    simulation A0+GB2, at $t=24\Tbar$. The solid black line represents
    the axisymmetric initial conditions.}\label{fig:hist_thick}
\end{figure}

Fig. \ref{fig:thick_panel} also confirms that the bar effects are
systematic and vary with the location of the volumes in the Galaxy.
We see that the main structures associated with the OLR Resonance
(Horn, separation between LSR and OLR Mode), are still present as
deformations in the 3rd, 4th and 5th contours of the distributions.
Fig. \ref{fig:neg_vR_thick} (equivalent to Fig. \ref{fig:neg_vR} for
the thin disk) indicates that, also in this case, the $v_R$ of the OLR
Mode varies significantly, as a function of $\phi$. However, in the
cases of volumes aligned with the bar ($\phi=0$ and
$\phi=90^{\circ}$), where $\langle v_R \rangle \sim 0$ as in the
axisymmetric case, the imprints of the bar on the $v_\phi$
distribution are not so clear as in the thin disk case, as
Fig. \ref{fig:hist_thick} shows.
\begin{figure}
  \centering
  \includegraphics[width=\columnwidth]{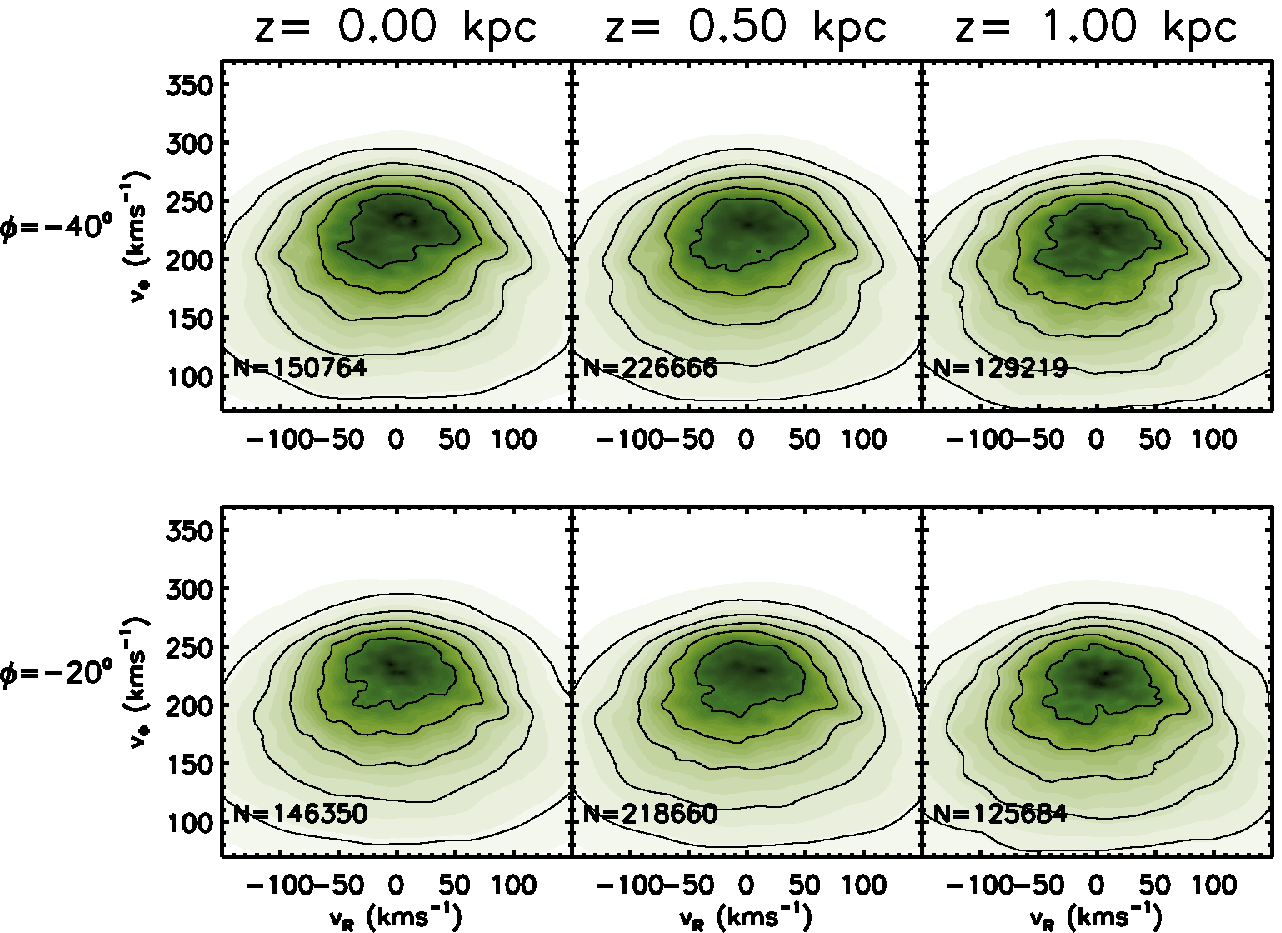}
  \caption{Kinematics of the thick disk simulation in the potential
    A0+GB2 in volumes centered at $R=8\Kpc$ and $\phi=-20^{\circ}$ and
    $\phi=-40^{\circ}$ and $z=0,0.5,1\Kpc$ (first, second and third
    column respectively). Notice that for $z>0$, we can double the
    resolution using the volumes centered at
    $-|z|$.}\label{fig:evz_thick}
\end{figure}

The vertical coherence of the kinematic structures in the thick disk
is depicted in Fig. \ref{fig:evz_thick}.  The vertical extent of the
thick disk allows us to have enough stars, even at $z=1\Kpc$.  The
distributions are similar on the plane and for $z=0.5\Kpc$ and
$z=1\Kpc$.  The Horn and the OLR Mode are present at the two angles,
visible especially in the 2nd, 3rd and 4th contours.  This is
confirmed by the analysis of the mean $v_R$ of the OLR Mode (shown in
the lower panels of Fig. \ref{fig:neg_vR_thick}).  Even for $z=2\Kpc$,
the periodicity of the mean $v_R$ of the OLR Mode is still noticeable.

We have repeated these analyses on the OLR Mode, both for the bar
models GB1 and LB2, and found that the same periodicity can be
observed, with a larger amplitude of the oscillations in the LB2 case.

\subsection{Combination of thin and thick disk}
\begin{figure}
  \centering
  \includegraphics[width=\columnwidth]{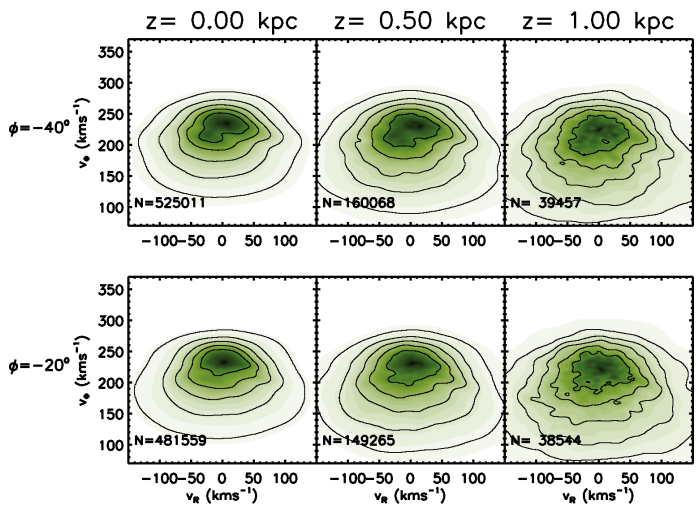}
  \caption{Kinematics of the thin and thick disk simulations together
    in the potential A0+GB2, in volumes centered at $R=8\Kpc$ and
    $\phi=-20^{\circ}$ and $\phi=-40^{\circ}$ and $z=0,0.5,1\Kpc$
    (first, second and third column respectively). This is obtained
    combining $5\times10^7$ thin disk particles with a a random
    subsample of $10^7$ thick disk particles, so that the
    thick-to-thin density normalization at $z=0$ is
    $\sim10\%$.}\label{fig:evz_both}
\end{figure}

In practice, surveys generally have a mix of both the thin and thick
disk populations. This is why in Fig. \ref{fig:evz_both} we show the
velocity distribution of the thin and the thick disk together.  Since
the thick disk has $20\%$ of the mass of the thin disk (Section
\ref{sect:axi}), we combine the $5\times10^7$ particles of the thin
disk for the A0+GB2 potential, with a random subsample of $10^7$ thick
disk particles run in the same potential.  As a result, in the volumes
on the Galactic plane the number of stars of the thick disk is $\sim
10\%$ that of the thin disk.  Fig. \ref{fig:evz_both} shows how, in a
sample of stars selected independently of the parent disk population,
the imprints of the dynamics of the bar are easily detectable. In
particular, we see how for $z=0$ the distribution completely resembles
that of the thin disk, while the thick disk, dominates the
distribution at $z\ge1\Kpc$. All the main features, described in the
previous figures (Horn, OLR Mode, LSR Mode) are easily detectable at
every height.

\section{Summary and discussion}\label{sect:con}

In this work we have studied the 3D response of test particles,
representing the thin and thick disk component of the Milky Way, to
the Galactic gravitational field including a rotating central bar.
Our goal was to understand if this non-axisymmetric component could
induce moving groups, especially in the case when the finite vertical
thickness of the Galaxy is taken into account.

We have found that the imprints of the bar are apparent both in a
population resembling the intermediate/old population of the thin disk
of the Milky Way and in a hot and vertically extended population
resembling the thick disk. Thanks to our 3D simulations, this is the
first time that such imprints have been traced far from the Galactic
plane, up to $z\sim 1\Kpc$ in the thin disk and up to $z\sim 2\Kpc$ in
the thick disk. These results are not strongly dependent on the bar
models used, as all the simulations explored with different structural
parameters (semi-major axes, vertical thickness and masses) yield
similar results.

The effects of the bar that are seen in our simulations are clearly
related to the resonant interaction between the rotation of the bar
and the orbits of the stars in the disk. Many of the stellar particles
in the vicinity of the Sun have orbits strongly affected by the bar's
Outer Lindblad Resonance of the bar. This OLR is apparent in a
splitting of the simulated velocity distributions in two main groups.
Our simulations also show that the impact of the Outer Lindblad
Resonance varies with position in the Galaxy, depending on
Galactocentric radius and angle from the major axis of the bar. On a
larger scale, the characteristics of the velocity distributions are
periodic with respect to the orientation angle of the bar, tracing the
symmetries of the bar's potential. This is manifested for instance, on
the vertex deviation around the Solar radius and in the mean radial
velocity of the Outer Lindblad Resonance groups.

In this sense our work agrees with previous results on the effects of
the Outer Lindblad Resonance of the bar on the kinematics of stars in
the Solar Neighbourhood (\citealt{Dehnen2000, Fux2001, Muhlbauer2003,
  GardnerFlynn2010}). The main difference is that our simulations are
more realistic as they are 3D and incorporate both a thin and a thick
disc.  For the analyses we have considered volumes of finite spatial
extent (as in the observations) in contrast to the previous studies,
which studied velocity distributions in a single point in the
configuration space. Due to this difference as well as our longer
integration times, we find more diffuse kinematic structures.

The presence of structures at large heights above the plane may also
be understood from the fact that near the Sun the forces along the
polar radius direction do not vary strongly with height, and certainly
much less than the vertical forces.  This implies that resonant
features in the $v_R-v_\phi$ velocity plane, even at large heights
from the plane, are expected. Our 3D simulations have shown that even
in these regions of the Galaxy where a large fraction of stars have
horizontal and vertical motions that are not decoupled, and the
orbital eccentricity is large (especially for the thick disk), the bar
significantly transform the velocity distribution, in a similar way as
on the plane.

We detect strong transient effects for approximately 10 bar rotation
periods after the bar is introduced. Short integration times, like in
e.g., \cite{Dehnen2000}, produce much clearer bar signatures. However,
we believe that it is unlikely that an old population would have
experienced the bar for such a short time scale.

Our simulations offer an %{\color{red} theoretical framework}
explanation for the discovery that several of the observed
kinematic groups in the Solar Neighbourhood are also present far below
the plane of Galaxy ($z\sim-0.7\Kpc$, \citealt{Antoja2012}).  If some
of these kinematic groups are direct consequence of the bar
gravitational force (like, e.g., the Hercules group seems to be), our
results show that it is not surprising to be able to recognize them
far from the Galactic plane.  Moreover, our simulations predict that,
depending on the location of the Sun in the Galaxy and with the
respect of the bar, some imprints of the bar should be recognizable
even beyond these heights.

Although we find kinematic structures in the thick disk, the Arcturus
stream does not appear in our simulations for our default (long)
integration times. This moving group is expected at $v_\phi \sim
134\kmsec$ in the Solar Neighbourhood of our simulations ($V\sim
100\kmsec$ in \citealt{Williams2009}).  However, for short integration
times and the long bar model, we can identify the same structures and
distortions in the low $v_\phi$ tail of the kinematic distribution in
our thin disk, already associated to Arcturus by
\cite{GardnerFlynn2010}.  Our simulations seem to suggest that, if the
Arcturus stream is an imprint of the bar, the bar should be very
young.  However, the lack of overdensity in the Arcturus region could
also be a reflection of the choice of the initial conditions of our
simulations if we are to associate it to a non-axisymmetry of the
potential rather than an accreted origin.

One important assumption of this work is that the the pattern speed
(and other properties) of the bar has not changed in the last $\sim 3
\Gyr$.  A change in the pattern speed should result in a change in the
resonances, and in which stars are affected.  But whether this
evolution would produce more or less bar signatures on the disc
phase-space remains to be seen.

Finally, as shown in \cite{Solway2012}, the spiral arms induce angular
momentum changes that affect not only the thin but also the thick
disc.  This suggests that spiral arms may also produce kinematic
signatures on the thick disc like those induced by the bar and studied
here.  It would be interesting to quantify the relative importance of
the effects of spiral arm resonances with respect to the bar's.

In our upcoming paper we will present a more exhaustive quantification
of the substructures and the bar effects and a dynamical
interpretation of the results based on the orbital frequency analysis.

The results obtained in this study unveil a more complex and structure
rich thick disk, which has been likely affected not only by external
accretion events but also by secular evolution induced by the disk
non-axisymmetries such as the bar and the spiral arms.

\begin{acknowledgements}
  The authors gratefully acknowledge support from the European
  Research Council under ERC Starting Grant GALACTICA-240271.
\end{acknowledgements}

\bibliography{paper1bib}{} 
\bibliographystyle{aa}

\end{document}